\documentclass[prd,aps,nofootinbib,showpacs,10pt]{revtex4}
\usepackage{amsmath}
\usepackage{amsmath,graphicx,color,epsfig}

\begin{document}
\title{Branching Ratios, Forward-backward Asymmetry and Angular Distributions of $B\to K_1l^+l^-$ Decays}
\author{ Run-Hui Li$^{a,b}$, Cai-Dian L\"u$^{a,c}$ and  Wei Wang$^{a}$\footnote{wwang@mail.ihep.ac.cn} }

\affiliation{
 \it $^a$ Institute of High Energy Physics, P.O. Box 918(4) Beijing 100049, People's Republic of China\\
 \it $^b$  School of Physics, Shandong University, Jinan 250100, People's Republic of China \\
 \it $^c$ Theoretical Physics Center for Science Facilities, Beijing 100049, People's Republic of China}

\begin{abstract}
Using the $B\to K_1$ form factors evaluated in the perturbative QCD
approach, we study semileptonic $B\to K_1(1270)l^+l^-$ and $B\to
K_1(1400)l^+l^-$ decays, where $K_1(1270)$ and $K_1(1400)$ are
mixtures of $K_{1A}$ and $K_{1B}$ which are $^3P_1$ and $^1P_1$
states, respectively. Using the technique of helicity amplitudes, we
express the decay amplitudes in terms of several independent and
Lorentz invariant pieces. We study the dilepton invariant mass
distributions, branching ratios, polarizations and forward-backward
asymmetries of $ B\to  K_1l^+l^-$ decays. The ambiguity in the sign
of the mixing angle will induce very large differences to branching
ratios of semileptonic $B$ decays: branching ratios without resonant
contributions either have the order of $10^{-6}$ or $10^{-8}$. But
the polarizations and the forward-backward asymmetries are not
sensitive to the mixing angles. We find that the resonant
contributions will dramatically change the dilepton invariant mass
distributions in the resonant region. We also provide the angular
distributions of $ B\to K_1l^+l^-\to (K\pi\pi)l^+l^-$ decays.
\end{abstract}

\maketitle

\section{Introduction}\label{section:introduction}

Semileptonic and nonleptonic $B$ decays provide an ideal place to
precisely test the standard model (SM) and search for new physics
(NP) scenarios beyond the SM. The flavor-changing
neutral-current(FCNC) $b\to s$ transitions are induced by the loop
effect in the SM, thus the relevant decay rates are typically small.
The NP models can either enhance the Wilson coefficients or
introduce new kinds of effective operators. Through the measurements
of observables such as branching ratios and direct CP asymmetries in
semileptonic and nonleptonic $B$ decays, experimentalists may figure
out those contributions from the NP models. Correspondingly, the
$b\to s$ transitions have received lots of consideration.

Besides the $B\to K^*\gamma$ and $B\to K^* l^+l^-$,  radiative and
semileptonic $B$ decays involving an axial-vector meson in the final
state are also sensitive to the NP contributions.  In the quark
model, there are two kinds of axial-vector mesons: the quantum
numbers are $J^{PC}=1^{++}$ or $J^{PC}=1^{+-}$, where $P,C$ denotes
the parity and charge parity of the meson, respectively. The two
strange axial-vector meons $K_{1A}$ and $K_{1B}$ can mix with each
other and form two physical states $K_1(1270)$ and $K_1(1400)$. But
the mixing angle is not uniquely determined only using the $\tau\to
K_1(1270)\bar\nu$ and $\tau\to K_1(1400)\bar\nu$ data. Thus
radiative and semileptonic $B$ decay channels  not only provide
valuable information on weak interactions but also offer an
opportunity to detect the internal structure of the $K_1$
mesons~\cite{Chen:2005cx,Hatanaka:2008xj,Aslam:2006vh,JamilAslam:2005mc,
Bashiry:2009wh,Bashiry:2009wq,Bayar:2008ug,Choudhury:2007kx,Paracha:2007yx,
Ahmed:2008ti,Saddique:2008xj,Hatanaka:2008gu,Sirvanli:2008nn}.

In our previous work~\cite{Wang:2007an,Li:2009tx}, we have studied
the $B\to A$ form factors in the perturbative QCD (PQCD) approach.
The predicted form factors are consistent with many nonleptonic
$B\to AP$ decays such as $\bar B^0\to a_1^\pm
\pi^\mp$~\cite{BtoAP1,BtoAP2}. Semileptonic $B\to Al\bar\nu$ decays
have also been studied and will be put on stringent experimental
tests in the future. In the present work, we will investigate the
$B\to K_1(1270)l^+l^-$ and $B\to K_1(1400)l^+l^-$ decays, including
the dilepton invariant mass distributions, branching ratios,
polarizations and forward-backward asymmetries. We will use the
helicity amplitudes to decompose the decay amplitudes into several
independent and Lorentz invariant pieces. All of them can be easily
evaluated in convenient frames. Our results are helpful to fix the
mixing angle with the help of the future data, for example,
branching ratios  are found to be sensitive to the sign of mixing
angles. Since the $K_1$ meson can only be reconstructed by at least
three pseudoscalar mesons, the angular distributions in $\bar B\to
\bar K_1l^+l^-\to (K\pi\pi)l^+l^-$ decays are investigated.

This paper is organized as follows. In Sec.
\ref{sec:effectiveHamitlonian}, we will introduce the effective
Hamiltonian which is responsible for $\bar B\to \bar K_1l^+l^-$
decays. Results for the $B\to K_1$ form factors in the perturbative
QCD approach are collected in Sec.~\ref{sec:formfactors}. In Sec.
\ref{sec:BtoK1lldecays}, we will construct the decay amplitudes in
terms of helicity amplitudes. The dilepton spectrum, the branching
ratios, and the forward-backward asymmetries in the $\bar B\to \bar
K_1l^+l^-$ decay are then studied. At the end of this section, we
present some discussions on angular distributions. Our conclusion is
given in the last section. In Appendices~\ref{appendix:C9eff},
\ref{appendix:mixing} and \ref{appendix:amplitudes}, we give the
explicit expressions for the effective Wilson coefficient
$C_9^{eff}$, the mixing of the $K_1$ mesons, and the technique of
helicity amplitudes, respectively.
\section{Effective Hamiltonian} \label{sec:effectiveHamitlonian}

The  effective Hamiltonian responsible for the  $b\to sl^+l^-$
transition is given by
 \begin{eqnarray}
 {\cal
 H}_{\rm{eff}}=-\frac{G_F}{\sqrt{2}}V_{tb}V^*_{ts}\sum_{i=1}^{10}C_i(\mu)O_i(\mu),\label{eq:Hamiltonian}
 \end{eqnarray}
where $V_{tb}$ and $V_{ts}$ are the Cabibbo-Kobayashi-Maskawa matrix
elements. $C_i(\mu)$ are the Wilson coefficients, and the local
operators $O_i(\mu)$ are given by~\cite{Buchalla:1995vs}
 \begin{eqnarray}
 O_1&=&(\bar s_{\alpha}c_{\alpha})_{V-A}(\bar
 c_{\beta}b_{\beta})_{V-A},\;\;
 O_2=(\bar
 s_{\alpha}c_{\beta})_{V-A}(\bar
 c_{\beta}b_{\alpha})_{V-A},\nonumber\\
 O_3&=&(\bar s_{\alpha}b_{\alpha})_{V-A}\sum_q(\bar
 q_{\beta}q_{\beta})_{V-A},\;\;
 O_4=(\bar s_{\alpha}b_{\beta})_{V-A}\sum_q(\bar
 q_{\beta}q_{\alpha})_{V-A},\nonumber\\
 O_5&=&(\bar s_{\alpha}b_{\alpha})_{V-A}\sum_q(\bar
 q_{\beta}q_{\beta})_{V+A},\;\;
 O_6=(\bar s_{\alpha}b_{\beta})_{V-A}\sum_q(\bar
 q_{\beta}q_{\alpha})_{V+A},\nonumber\\
 O_7&=&\frac{e m_b}{8\pi^2}\bar
 s\sigma^{\mu\nu}(1+\gamma_5)bF_{\mu\nu}+\frac{e m_s}{8\pi^2}\bar
 s\sigma^{\mu\nu}(1-\gamma_5)bF_{\mu\nu},\nonumber\\
 O_9&=&\frac{\alpha_{\rm{em}}}{2\pi}(\bar l\gamma_{\mu}l)(\bar
 s\gamma^{\mu}(1-\gamma_5)b),\;\;
 O_{10}=\frac{\alpha_{\rm{em}}}{2\pi}(\bar l\gamma_{\mu}\gamma_5l)(\bar
 s\gamma^{\mu}(1-\gamma_5)b),\label{eq:operators}
 \end{eqnarray}
where $(\bar q_1q_2)_{V-A}(\bar q_3 q_4)_{V-A}\equiv(\bar q_1
\gamma^{\mu}(1-\gamma_5)q_2)(\bar q_3\gamma_{\mu}(1-\gamma)q_4)$,
and $(\bar q_1q_2)_{V-A}(\bar q_3 q_4)_{V+A}\equiv(\bar q_1
\gamma^{\mu}(1-\gamma_5)q_2)(\bar q_3\gamma_{\mu}(1+\gamma)q_4)$.
The amplitude for $b\to sl^+l^-$ transition can be decomposed as
 \begin{eqnarray}
 {\cal A}(b\to sl^+
 l^-)&=&\frac{G_F}{2\sqrt{2}}\frac{\alpha_{\rm{em}}}{\pi}V_{tb}V^*_{ts}\bigg\{
 C_9^{\rm{eff}}(q^2)
 [\bar s \gamma_{\mu}(1-\gamma_5)b][\bar l\gamma^{\mu}l] + C_{10}[\bar s\gamma_{\mu}(1-\gamma_5)b]
 [\bar l\gamma^{\mu}\gamma_5l]\nonumber\\
 &&- 2m_bC_7^{\rm{eff}}\big[\bar s i\sigma_{\mu\nu}\frac{q^{\nu}}{q^2}
 (1+\gamma_5)b\big][\bar l\gamma^{\mu}l]- 2m_sC_7^{\rm{eff}}\big[\bar s i\sigma_{\mu\nu}\frac{q^{\nu}}{q^2}
 (1-\gamma_5)b\big ][\bar l\gamma^{\mu}l] \bigg\},\label{eq:Ampbtos}
 \end{eqnarray}
with $m_b$ as the $b$ quark mass in the $\overline{\mbox{MS}}$
scheme. $C_7^{\rm{eff}}=C_7-C_5/3-C_6$ and $C_9^{\rm{eff}}$ contains
both the long-distance and short-distance contributions, which is
given by
 \begin{eqnarray}
 C_9^{\rm{eff}}(q^2)&=&C_9(\mu)+Y_{\rm{pert}}(\hat{s})+Y_{\rm{LD}}(q^2).
 \end{eqnarray}
with $\hat{s}=q^2/m_B^2$. $Y_{\rm{pert}}$ represents the
perturbative contributions, and $Y_{\rm{LD}}$ is the long-distance
part. For the explicit expressions of $Y_{\rm{pert}}$ and
$Y_{\rm{LD}}$, see Appendix~\ref{appendix:C9eff}.

\section{$B\to K_1$ Form factors}\label{sec:formfactors}

$\bar B\to A$ transition form factors are defined by
 \begin{eqnarray}
  \langle A(P_2,\epsilon^*)|\bar q\gamma^{\mu}\gamma_5 b|\bar B(P_B)\rangle
   &=&-\frac{2iA(q^2)}{m_B-m_A}\epsilon^{\mu\nu\rho\sigma}
     \epsilon^*_{\nu}P_{B\rho}P_{2\sigma}, \nonumber\\
  \langle A(P_2,\epsilon^*)|\bar q\gamma^{\mu}b|\bar
  B(P_B)\rangle
   &=&-2m_A V_0(q^2)\frac{\epsilon^*\cdot q}{q^2}q^{\mu}
    -(m_B-m_A)V_1(q^2)\left[\epsilon^*_{\mu}
    -\frac{\epsilon^*\cdot q}{q^2}q^{\mu} \right] \nonumber\\
    &&+V_2(q^2)\frac{\epsilon^*\cdot q}{m_B-m_A}
     \left[ (P_B+P_2)^{\mu}-\frac{m_B^2-m_A^2}{q^2}q^{\mu} \right],\nonumber\\
  \langle A(P_2,\epsilon^*)|\bar q\sigma^{\mu\nu}\gamma_5q_{\nu}b|\bar
  B(P_B)\rangle
   &=&-2T_1(q^2)\epsilon^{\mu\nu\rho\sigma}
     \epsilon^*_{\nu}P_{B\rho}P_{2\sigma}, \nonumber\\
  \langle A(P_2,\epsilon^*)|\bar q\sigma^{\mu\nu}q_{\nu}b|\bar
  B(P_B)\rangle
   &=&-iT_2(q^2)\left[(m_B^2-m_A^2)\epsilon^{*\mu}
       -(\epsilon^*\cdot q)(P_B+P_2)^{\mu} \right]\nonumber\\
   &&-iT_3(q^2)(\epsilon^*\cdot q)\left[
       q^{\mu}-\frac{q^2}{m_B^2-m_A^2}(P_B+P_2)^{\mu}\right],\label{eq:BtoK1formfactor}
 \end{eqnarray}
where $m_A$ is the mass of the axial-vector meson.  The relation
$2m_AV_0=(m_B-m_A)V_1-(m_B+m_A)V_2$ is obtained at $q^2=0$. In the
PQCD approach, we find that this relation for the whole kinematic
region becomes
\begin{eqnarray}
 V_2&=&\Big[\frac{(1-r_2)^2}{\rho}V_1-\frac{2r_2(1+r_2)V_0}{\rho}\Big],
\end{eqnarray}
where $r_2=\frac{m_A}{m_B}$ and $\rho=1-q^2/m_B^2$.  Numerical
results for the $B\to K_1(1270)$ and $B\to K_1(1400)$ form factors
are quoted from our previous work~\cite{Wang:2007an,Li:2009tx}  and
collected in Table~\ref{Tab:formfactorsBtoAbeforemixing}. The form
factors in the large recoiling region are directly calculated. In
order to extrapolate the form factors to the whole kinematic region,
we have adopted the dipole parametrization for the form factors£º
\begin{eqnarray}
 F(q^2)=\frac{F(0)}{1-a(q^2/m_B^2)+b(q^2/m_B^2)^2}\;.
\end{eqnarray}
The errors in the results are from: the decay constant of  $B$ meson
and shape parameter $\omega_b$;
$\Lambda_{\rm{QCD}}\big((0.25\pm0.05)\rm{GeV}\big)$ and the
factorization scales; Gegenbauer moments of axial-vectors'
light-cone distribution amplitudes (LCDAs).

\begin{table}
\caption{$B\to K_{1A,1B}$ form factors. $a,b$ are the parameters of
the form factors in dipole parametrization. The errors are from:
decay constant of $B$ meson and shape parameter $\omega_b$;
$\Lambda_{\rm{QCD}}$ and the scales $t_e$; Gegenbauer moments of
axial-vectors' LCDAs. The three kinds of uncertainties for the
fitter parameters $a$ and $b$ are quadratically added together. }
 \label{Tab:formfactorsBtoAbeforemixing}
 \begin{center}
 \begin{tabular}{|c|c c c|c|c c c|}
\hline \hline
 $F$       & $F(0)$  & $a$ &$b$       & $F$       & $F(0)$  & $a$ &$b$                \\
 \hline
\hline
 $A^{B K_{1A}}$      &$0.27_{-0.05-0.01-0.06}^{+0.06+0.00+0.06}$    &$1.73_{-0.06}^{+0.07}$    &$0.67_{-0.07}^{+0.09}$  &$A^{B K_{1B}}$   &$0.20_{-0.04-0.01-0.05}^{+0.04+0.01+0.05}$    &$1.73_{-0.06}^{+0.07}$    &$0.68_{-0.06}^{+0.08}$   \\
\hline
 $V_0^{B K_{1A}}$    &$0.35_{-0.07-0.02-0.13}^{+0.08+0.01+0.13}$    &$1.73_{-0.09}^{+0.07}$    &$0.66_{-0.10}^{+0.09}$  &$V_0^{B K_{1B}}$   &$0.52_{-0.10-0.02-0.07}^{+0.12+0.01+0.07}$    &$1.72_{-0.06}^{+0.06}$    &$0.64_{-0.06}^{+0.07}$  \\
\hline
 $V_1^{B K_{1A}}$    &$0.47_{-0.09-0.01-0.01}^{+0.11+0.01+0.01}$    &$0.75_{-0.04}^{+0.09}$    &$-0.13_{-0.00}^{+0.10}$  &$V_1^{B K_{1B}}$   &$0.36_{-0.07-0.02-0.08}^{+0.08+0.01+0.09}$    &$0.78_{-0.05}^{+0.06}$    &$-0.10_{-0.03}^{+0.05}$  \\
\hline
  $V_2^{B K_{1A}}$   &$0.14_{-0.03-0.01-0.02}^{+0.03+0.00+0.02}$  &$--$  &$--$   &$V_2^{B K_{1B}}$    &$0.00_{-0.00-0.00-0.03}^{+0.00+0.00+0.03}$  &$--$  &$--$       \\
\hline
 $T_1^{B K_{1A}}$   &$0.37_{-0.07-0.01-0.01}^{+0.08+0.01+0.01}$    &$1.70_{-0.07}^{+0.08}$    &$0.63_{-0.09}^{+0.08}$  &$T_1^{B K_{1B}}$   &$0.29_{-0.06-0.01-0.06}^{+0.06+0.01+0.06}$     &$1.68_{-0.07}^{+0.08}$    &$0.61_{-0.06}^{+0.10}$  \\
\hline
 $T_2^{B K_{1A}}$   &$0.37_{-0.07-0.01-0.01}^{+0.08+0.01+0.01}$    &$0.72_{-0.07}^{+0.10}$    &$-0.16_{-0.01}^{+0.06}$  &$T_2^{B K_{1B}}$   &$0.29_{-0.06-0.01-0.06}^{+0.06+0.01+0.06}$      &$0.73_{-0.07}^{+0.07}$    &$-0.14_{-0.04}^{+0.03}$   \\
\hline
 $T_3^{B K_{1A}}$   &$0.33_{-0.07-0.01-0.08}^{+0.08+0.00+0.08}$    &$1.61_{-0.06}^{+0.09}$    &$0.54_{-0.05}^{+0.11}$  &$T_3^{B K_{1B}}$   &$0.20_{-0.04-0.01-0.05}^{+0.05+0.01+0.05}$      &$1.38_{-0.09}^{+0.08}$    &$0.43_{-0.07}^{+0.06}$  \\
 \hline \hline
 \end{tabular}
 \end{center}
 \end{table}

\section{Semileptonic  $B\to K_1(1270)l^+l^-$ and $B\to K_1(1400)l^+l^-$ decays}\label{sec:BtoK1lldecays}

Physical states $K_1(1270)$ and $K_1(1400)$ are mixtures of $K_{1B}$
and $K_{1A}$, while the mixing angle is usually constrained using
the $\tau^-$ decays. The solution for the mixing angle is found to
be twofold: $\theta_K=(-38\pm11)^\circ$ and
$\theta_K=(48.5\pm11.5)^\circ$. The large uncertainties mainly arise
from the experimental data of the branching ratios, see Appendix
\ref{appendix:mixing}. In the following, we will focus our
investigation on these two ranges of the mixing angle. By
reexpressing the metric tensor $g_{\mu\nu}$ with the polarization
vectors and momentum, we can decompose the amplitude of semileptonic
decays into two Lorentz invariant part, the leptonic amplitude
$L(L/R,i)$ and the hadronic amplitude $H(L/R,i)$, where $i=0,+$ or
$-$ denotes the three different polarizations of the axial-vector
$K_1$. For more details about the helicity amplitudes and
definitions of the angles in the angular distribution, see Appendix
\ref{appendix:amplitudes}.

Combining the leptonic amplitudes, the hadronic amplitudes and the
phase space together,  the partial decay width is given as
\begin{eqnarray}
 d\Gamma_i(\bar B\to \bar K_1l^+l^-)&=&\frac{\sqrt{\lambda}}{1024 \pi^4 m_B^3}
 d\cos\theta_1 d\phi dq^2 |{\cal A}_i(B\to
 K_1l^+l^-)|^2\nonumber\\
 &=&\frac{\sqrt{\lambda}}{1024 \pi^4 m_B^3}
 d\cos\theta_1 d\phi dq^2 (|{L}(L,i){H}(L,i)|^2+|{L}(R,i){H}(R,i)|^2),\label{eq:Pdecaywith}
\end{eqnarray}
where $\lambda=(m_A^2+m_B^2-q^2)^2-4m_B^2m_A^2 =
(m_B^2-m_A^2-q^2)^2-4m_A^2q^2$.
\subsection{Dilepton Mass Distributions}
 \begin{figure}
 \begin{center}
 \includegraphics[width=7.cm]{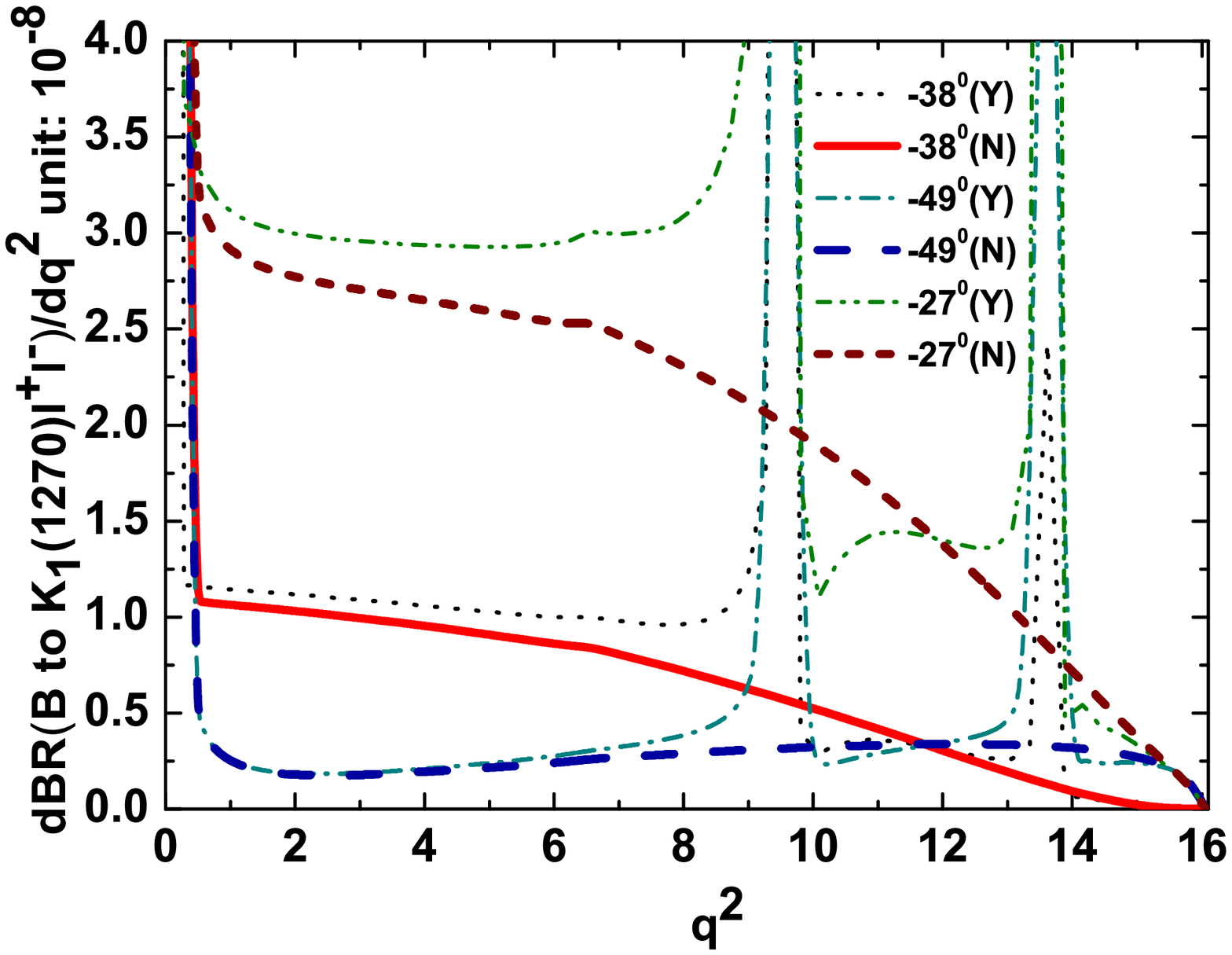}
 \includegraphics[width=7.cm]{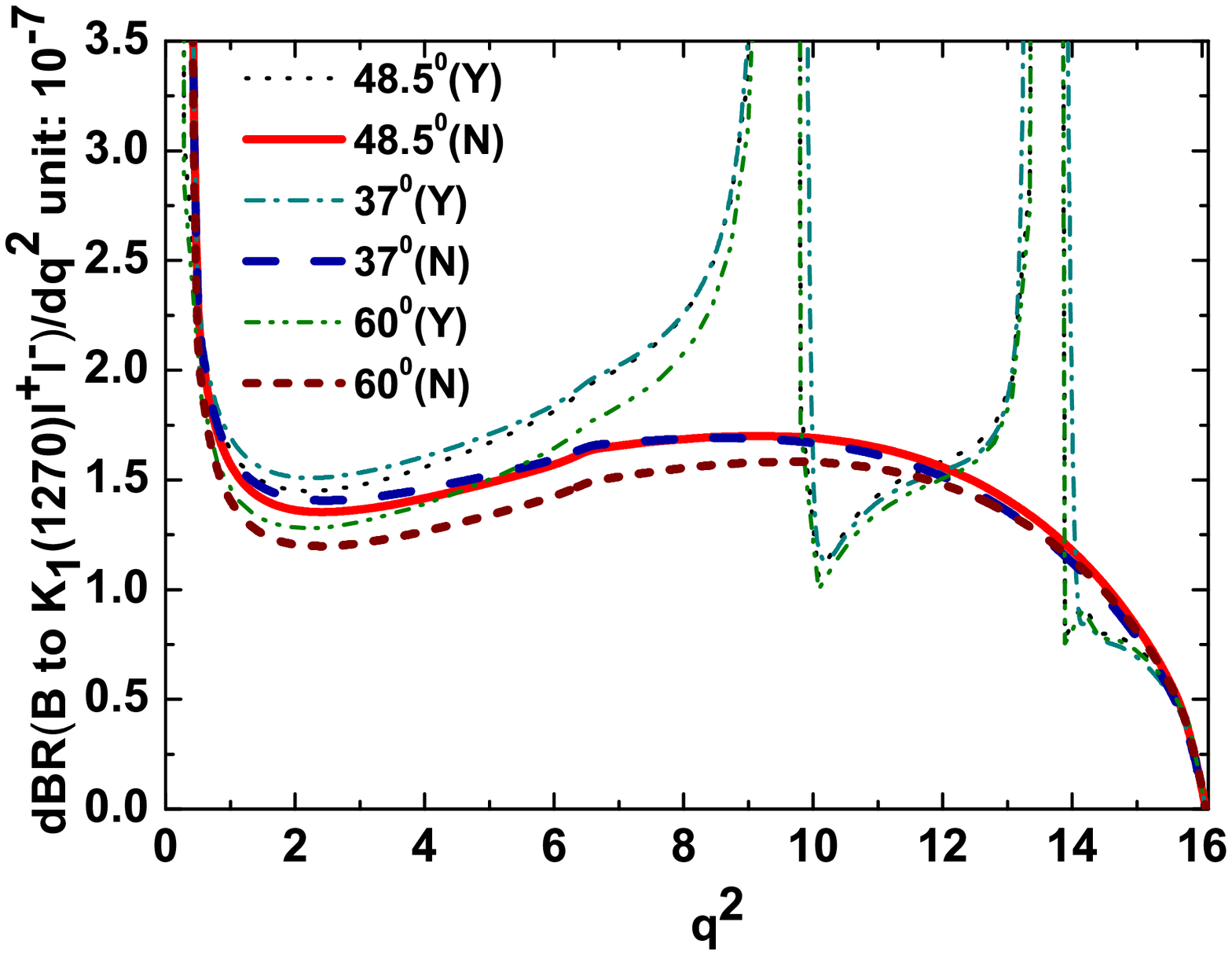}
 \includegraphics[width=7.cm]{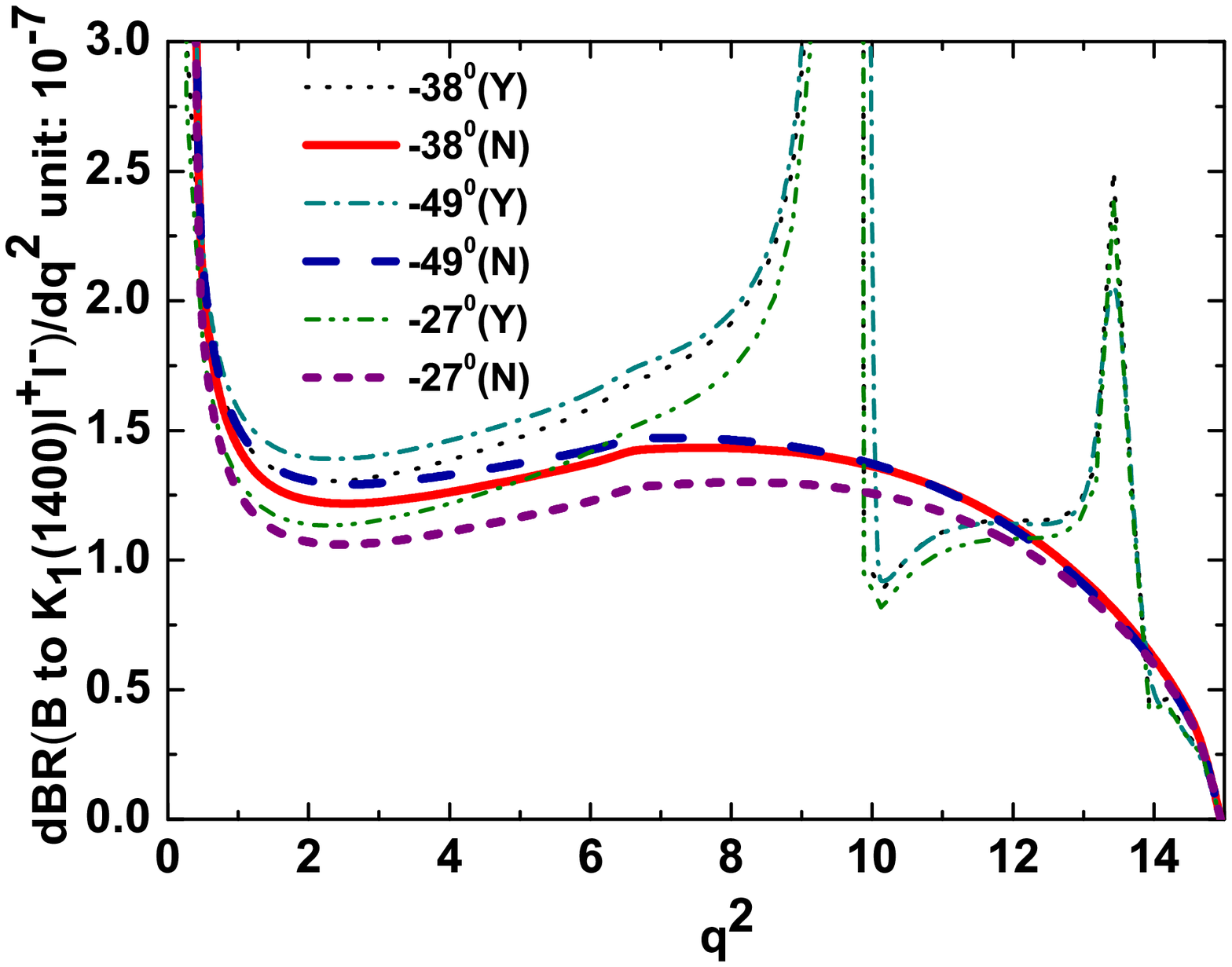}
 \includegraphics[width=7.cm]{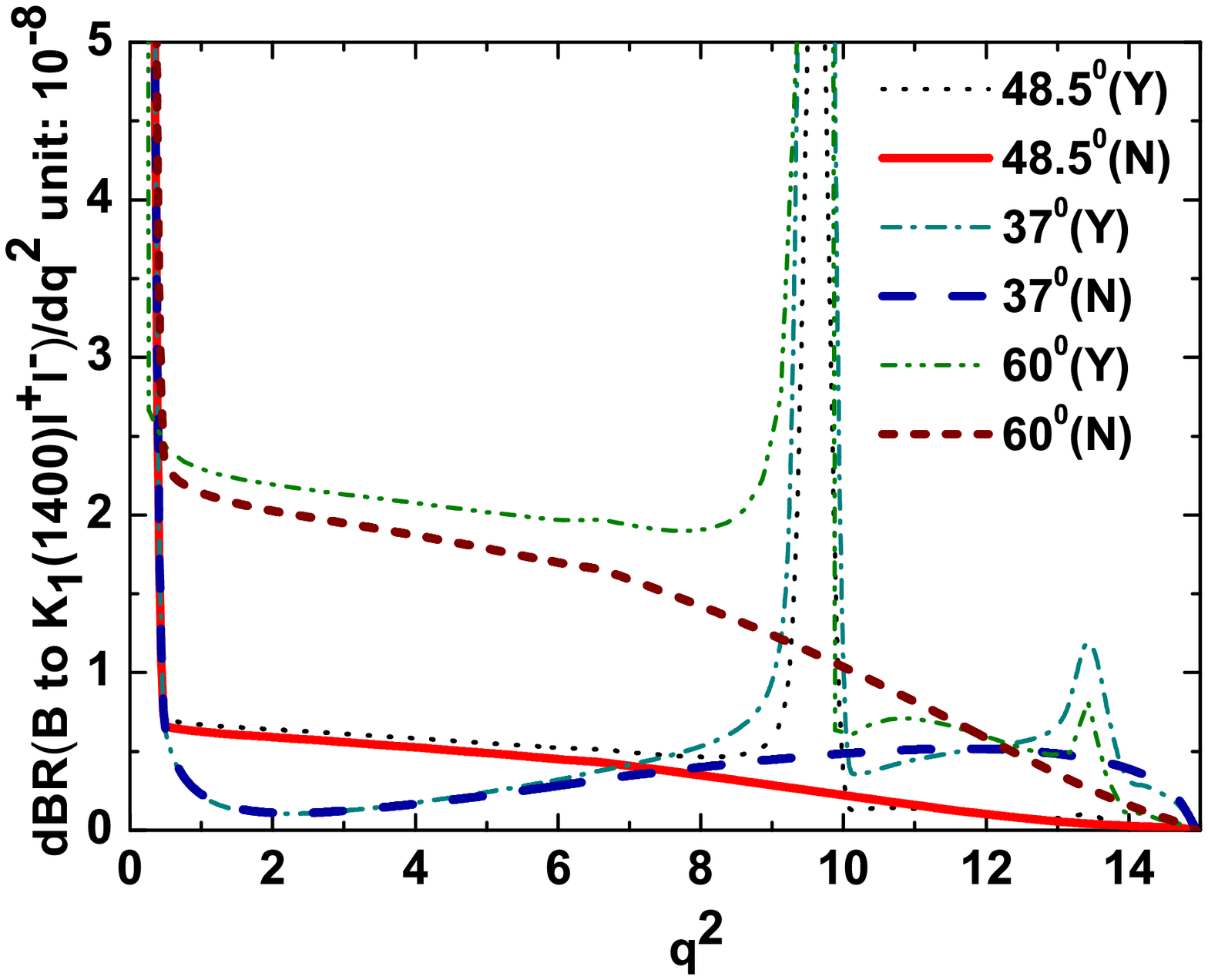}
\caption{The dependence on $q^2$ of $\frac{d{\cal BR}(B\to
K_1(1270)l^+l^-)}{dq^2}$ and $\frac{d{\cal BR}(B\to
K_1(1400)l^+l^-)}{dq^2}$. Since the differential decay rates depends
on the mixing angle, we have given the results at several mixing
angles as shown in the diagrams. The character "Y"("N") in the
brackets in the diagrams denotes that the resonant contributions are
(not) taken into account.}
 \label{fig:Brq2}
 \end{center}
 \end{figure}
With $\theta_1$ and $\phi$ integrated out in Eq.
(\ref{eq:Pdecaywith}), one obtains the dilepton mass spectrum of
$B\to K_1l^+l^-$ decay:
 \begin{eqnarray}
 \frac{d\Gamma_i}{dq^2}=\frac{\sqrt{\lambda}q^2}{96 \pi^3 m_B^3}
 \bigg[|H(L,i)|^2+|H(R,i)|^2\bigg].\label{eq:Gammaq2}
 \end{eqnarray}
The differential decay rates can be obtained by summing the three
different polarizations of the axial-vector meson:
 \begin{eqnarray}
 \frac{d{\cal BR}(B \to K_1l^+l^-)}{dq^2}=\sum_{i=0,+,-}\frac{d\Gamma_i}{dq^2}\tau_B,
 \end{eqnarray}
where $\tau_B$ is the lifetime of the $B$ meson.

In Fig. \ref{fig:Brq2}, we give our results of the differential
decay rates of $B^-\to K_1^-l^+l^-$. To show the dependence on
mixing angles, we give the results at several reference points:
$\theta_K=-27^{\circ}, -38^{\circ}, -49^{\circ}$ and
$\theta=37^{\circ}, 48.5^{\circ}, 60^{\circ}$.  The first feature in
these diagrams is the threshold enhancement for the dilepton's
invariant mass distributions.  These enhancements are caused by the
$q^2$ in the denominators of the terms with $C_{7L}$ and $C_{7R}$ in
the transverse decay widths as shown in  Eqs. (\ref{eq:HL+}),
(\ref{eq:HL-}), (\ref{eq:HR+}) and (\ref{eq:HR-}). If the leptons'
masses are taken into account, the invariant masses will have a
minimum $q^2_{\rm min}=4m_l^2$ and the enhancement is expected to
become much smoother. From Eq.~(\ref{eq:yld}), we know that the
resonant contributions can give enhancements to the partial decay
widths around the region $s\sim m_V^2$. For the $J/\Psi$,
$\Psi(2S)$, the diagram clearly shows the resonant contributions
from these two vector mesons. However, because of the small
branching fraction of $\Psi\to l^+l^-$ and the large decay width of
$\Psi$, resonant contributions from the other resonances are highly
suppressed and thus they are not very manifest in the diagrams.

\subsection{Decay Widths and Branching Ratios}

Integrating over $q^2$ in (\ref{eq:Gammaq2}), one can obtain the
decay width
 \begin{eqnarray}
 \Gamma_i(\bar B\to \bar K_1l^+l^-)&=&\int_{0.1 {\rm GeV}^2}^{(m_B-m_{K_1})^2} dq^2 \frac{\sqrt{\lambda}q^2}{96 \pi^3 m_B^3}
 \bigg[|H(L,i)|^2+|H(R,i)|^2\bigg],\\
 \Gamma(\bar B\to \bar K_1l^+l^-)&=&\sum_{i=0,\pm}\Gamma_i(\bar B\to \bar K_1l^+l^-)
 \end{eqnarray}
where we have introduced a small cutoff for the invariant mass of
the lepton pair in the integration. The three branching ratios are
given by
\begin{eqnarray}
 {\cal BR}_i=\frac{\Gamma_i}{\Gamma_{\rm tot}},
\end{eqnarray}
where $\Gamma_{\rm tot}$ is the decay width of the $B$ meson. The
transverse and total branching ratios are defined by
\begin{eqnarray}
 {\cal BR}_T&=&{\cal BR}_++{\cal BR}_-,\\
 {\cal BR}_{\rm tot}&=&{\cal BR}_0+{\cal BR}_T.
\end{eqnarray}
The polarization parameter, the ratio of the longitudinal and
transverse decay width, is defined by
\begin{eqnarray}
 R_{L/T}=\frac{\Gamma_0}{\Gamma_++\Gamma_-}=\frac{{\cal BR}_0}{{\cal
 BR}_T}.
\end{eqnarray}

\begin{table}
\begin{center}
\caption{The branching ratios of $\bar B\to \bar K_1l^+l^-$ decays
without resonant contributions(unit:$10^{-6}$).  ${\cal BR}_T={\cal
BR}_++ {\cal BR}_-$ and ${\cal BR}_{\rm tot}={\cal BR}_0 + {\cal
BR}_T$. For each channel, the results in the first line are obtained
with $\theta_K=-38^\circ$ and the second line with
$\theta_K=48.5^\circ$. The errors are from the $B\to K_1$ form
factors.}
 \label{tab:results_Br}
 \begin{tabular}{c c c c c c}
 \hline\hline
 \ \ \           &${\cal BR}_0$        &${\cal BR}_+$        &${\cal BR}_-$       &${\cal BR}_{\rm tot}$         &$R_{L/T}$    \\
 \hline
 \ \ \ $B\to K_1(1270)l^+l^-$   &$0.1_{-0.1}^{+0.2}$   &$<0.001$   &$<0.001$   &$0.1_{-0.1}^{+0.2}$   &$...$  \\
 \ \ \                          &$1.5_{-0.7}^{+0.8}$   &$0.03_{-0.01}^{+0.01}$   &$0.8_{-0.3}^{+0.4}$   &$2.3_{-1.0}^{+1.2}$   &$2.0\pm0.2$  \\
 \hline
 \ \ \ $B\to K_1(1400)l^+l^-$   &$1.2_{-0.5}^{+0.7}$   &$0.02_{-0.01}^{+0.01}$   &$0.6_{-0.2}^{+0.3}$   &$1.8_{-0.8}^{+1.0}$   &$2.0\pm0.2$  \\
 \ \ \                          &$0.05_{-0.06}^{+0.14}$   &$<0.001$   &$\sim 0.004$   &$0.05_{-0.03}^{+0.14}$   &...  \\
 \hline\hline
 \end{tabular}
 \end{center}
 \end{table}

Our predictions on the branching ratios and polarizations are
collected in Table \ref{tab:results_Br}, where the resonant
contributions from $V(\bar cc)$ are not taken into account.  From
this table, we can see that the total branching ratios are sensitive
to the mixing angles. The two kinds of $B\to K_1$ form factors shown
in Table~\ref{Tab:formfactorsBtoAbeforemixing} are similar in size
but have different signs. If the mixing angle is chosen as
$-38^\circ$ which is very close to $-45^\circ$, the $B\to K_1(1270)$
form factors are expected to be very small while the $B\to
K_1(1400)$ form factors are large. Thus the branching fraction of
$B\to K_1(1270)l^+l^-$ is much larger than that of $B\to
K_1(1400)l^+l^-$. On the contrary, if the mixing angle is chosen as
$48.5^\circ$ which is close to $45^\circ$, the ${\cal BR}(B\to
K_1(1270)l^+l^-)$ is much smaller than ${\cal BR}(B\to
K_1(1400)l^+l^-)$. It is clear that the three branching fractions
obey the relation for all mixing angles: ${\cal BR}_0(B\to
K_1l^+l^-)>{\cal BR}_-(B\to K_1l^+l^-)>{\cal BR}_+(B\to K_1l^+l^-)$.
From Table~\ref{Tab:formfactorsBtoAbeforemixing}, we can see the
form factors $T_1$ and $T_2$ have similar magnitudes at $q^2=0$, and
so are $V_1$ and $A$. Thus the functions $H(L,+)$ and $H(R,+)$ are
suppressed due to the destructive contributions from these form
factors while the other two functions $H(L,-)$ and $H(R,-)$ are
enhanced.

\begin{figure}
 \begin{center}
 \includegraphics[width=7.cm]{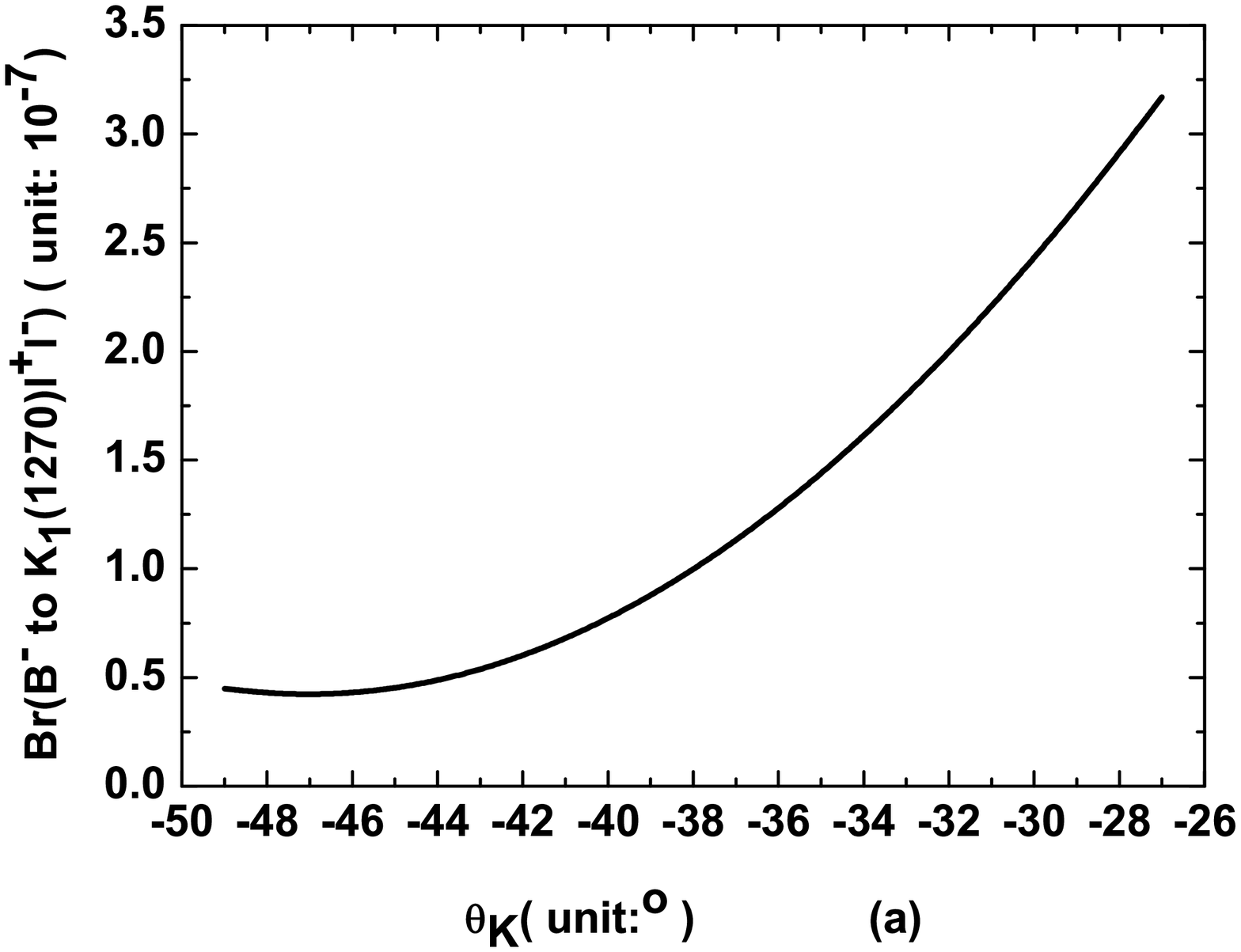}
 \includegraphics[width=7.cm]{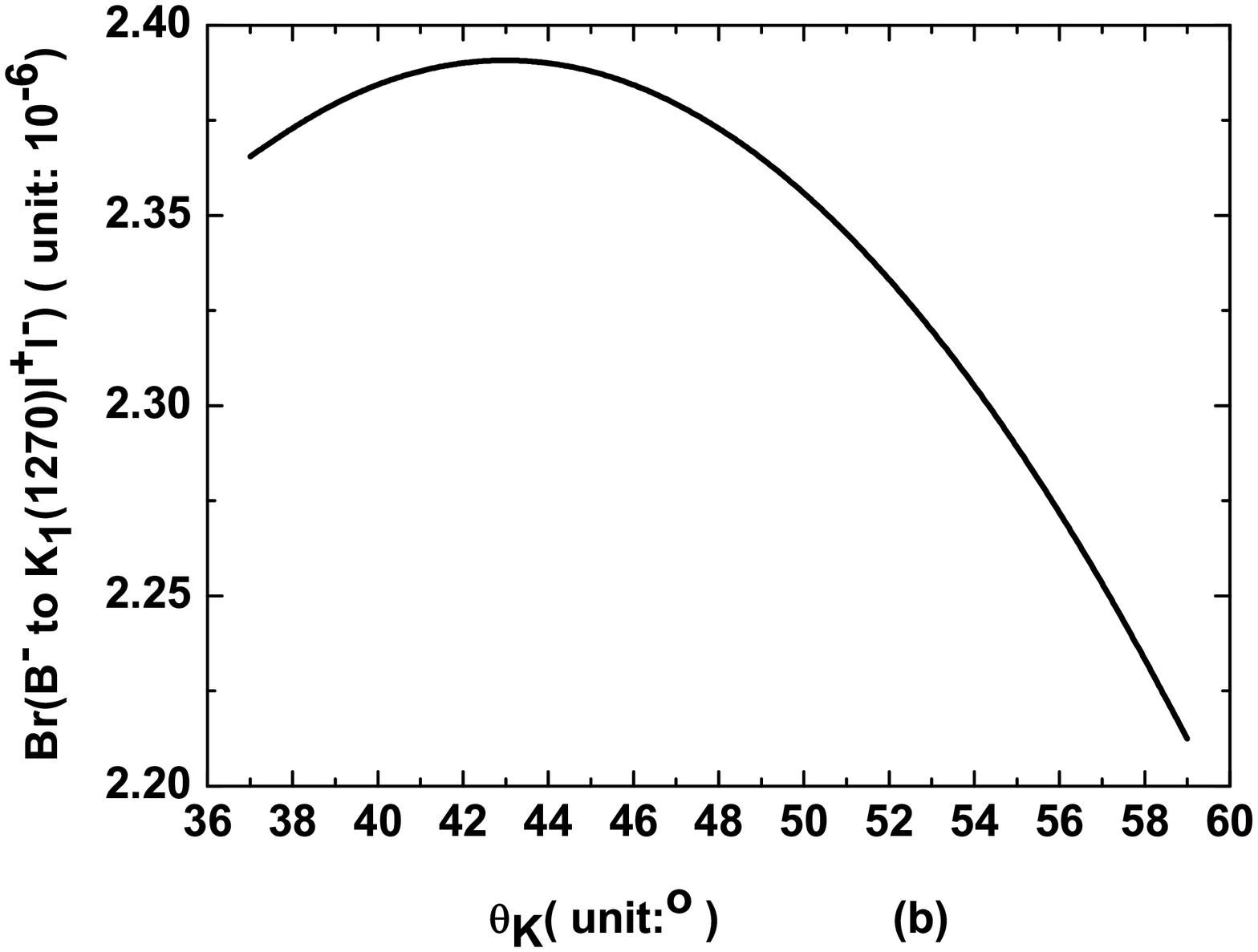}
 \includegraphics[width=7.cm]{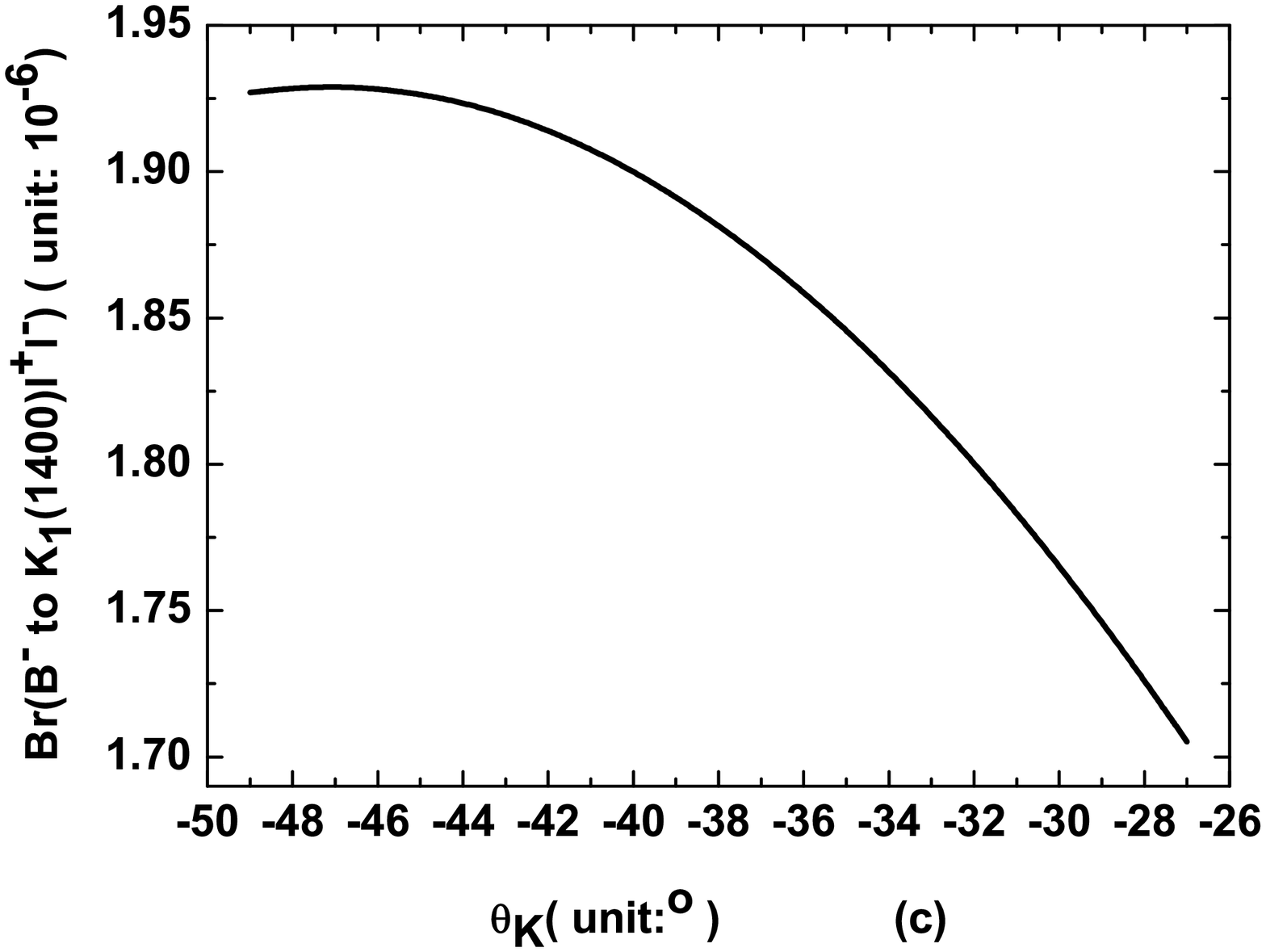}
 \includegraphics[width=7.cm]{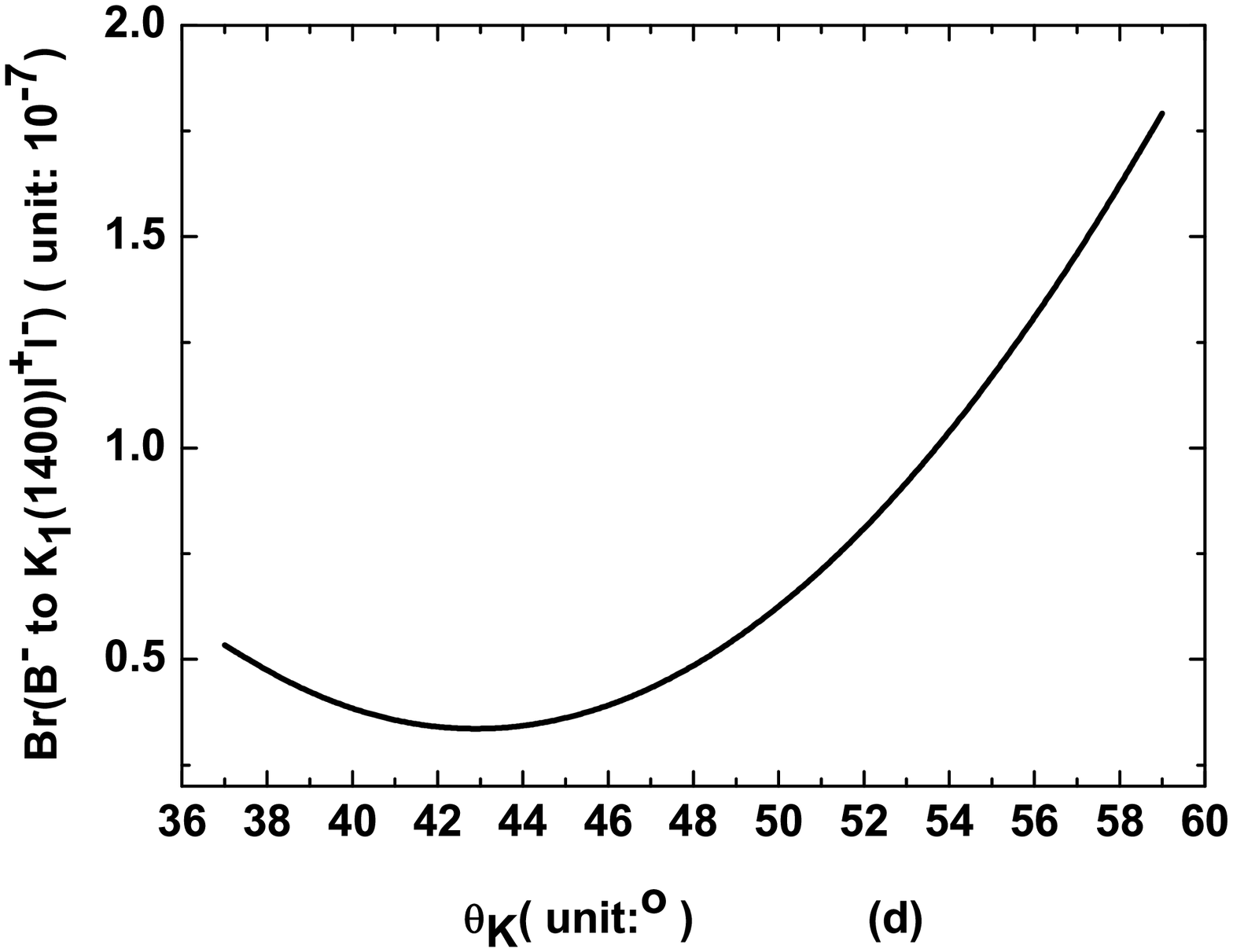}
 \caption{$\theta_K$ dependence of the branching ratios without resonances.}
 \label{fig:Brtheta}
 \end{center}
 \end{figure}

Since there are large uncertainties in the mixing angle $\theta_K$,
we give  the dependence on the mixing angle $\theta_K$ of the
branching ratios in Fig.~\ref{fig:Brtheta}. As indicated from these
diagrams, the ranges of the branching ratios shown in diagram (a),
(b), (c) and (d) are $4.2\times 10^{-8}\sim3.2\times 10^{-7}$,
$2.2\times 10^{-6}\sim2.4\times 10^{-6}$, $1.7\times 10^{-6}\sim
1.9\times 10^{-6}$, and $3.3\times 10^{-8} \sim 1.8\times 10^{-7}$,
respectively.

\subsection{Ratios of decay widths}
\begin{figure}
 \begin{center}
 \includegraphics[width=7.cm]{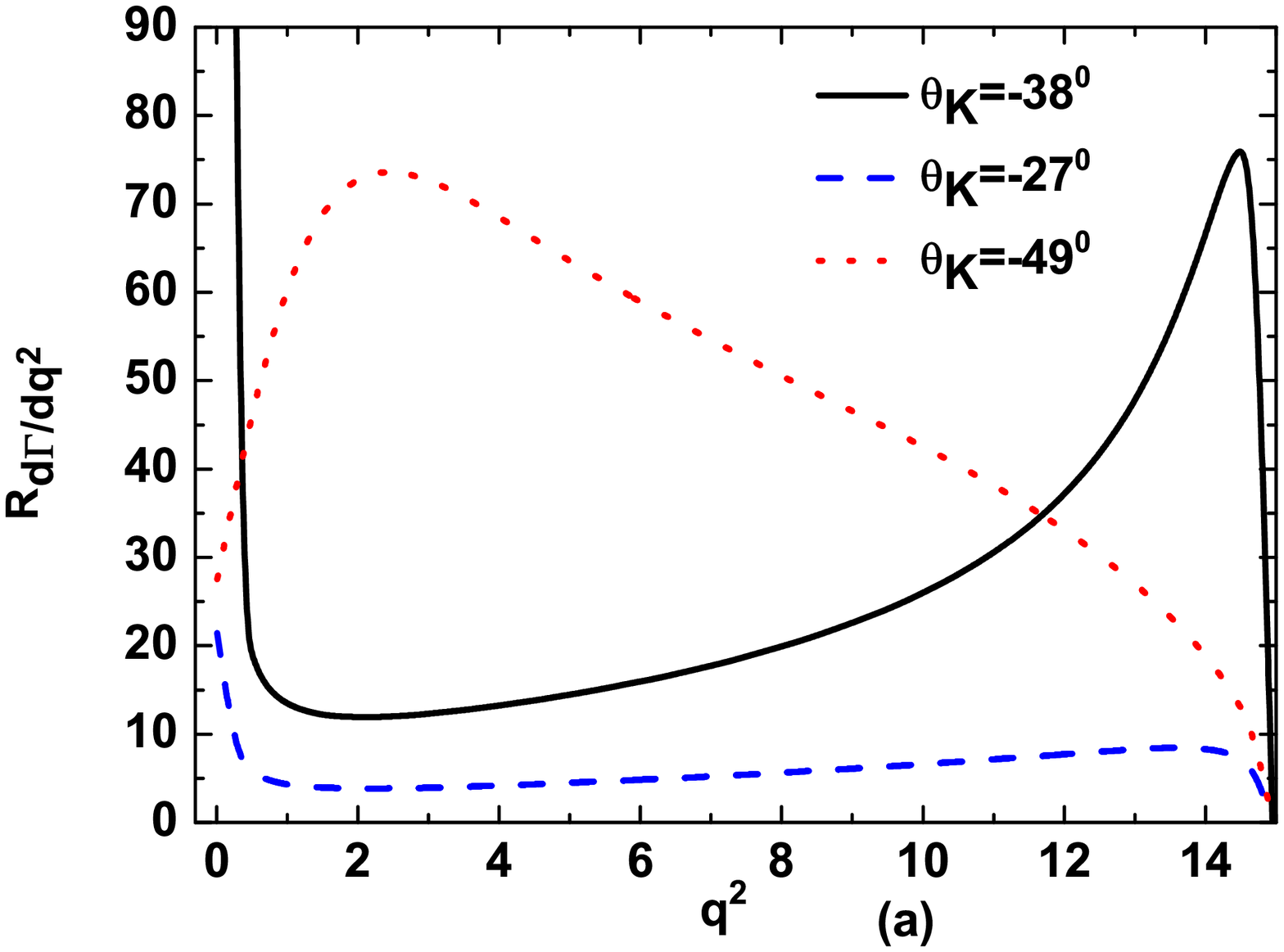}
 \includegraphics[width=7.cm]{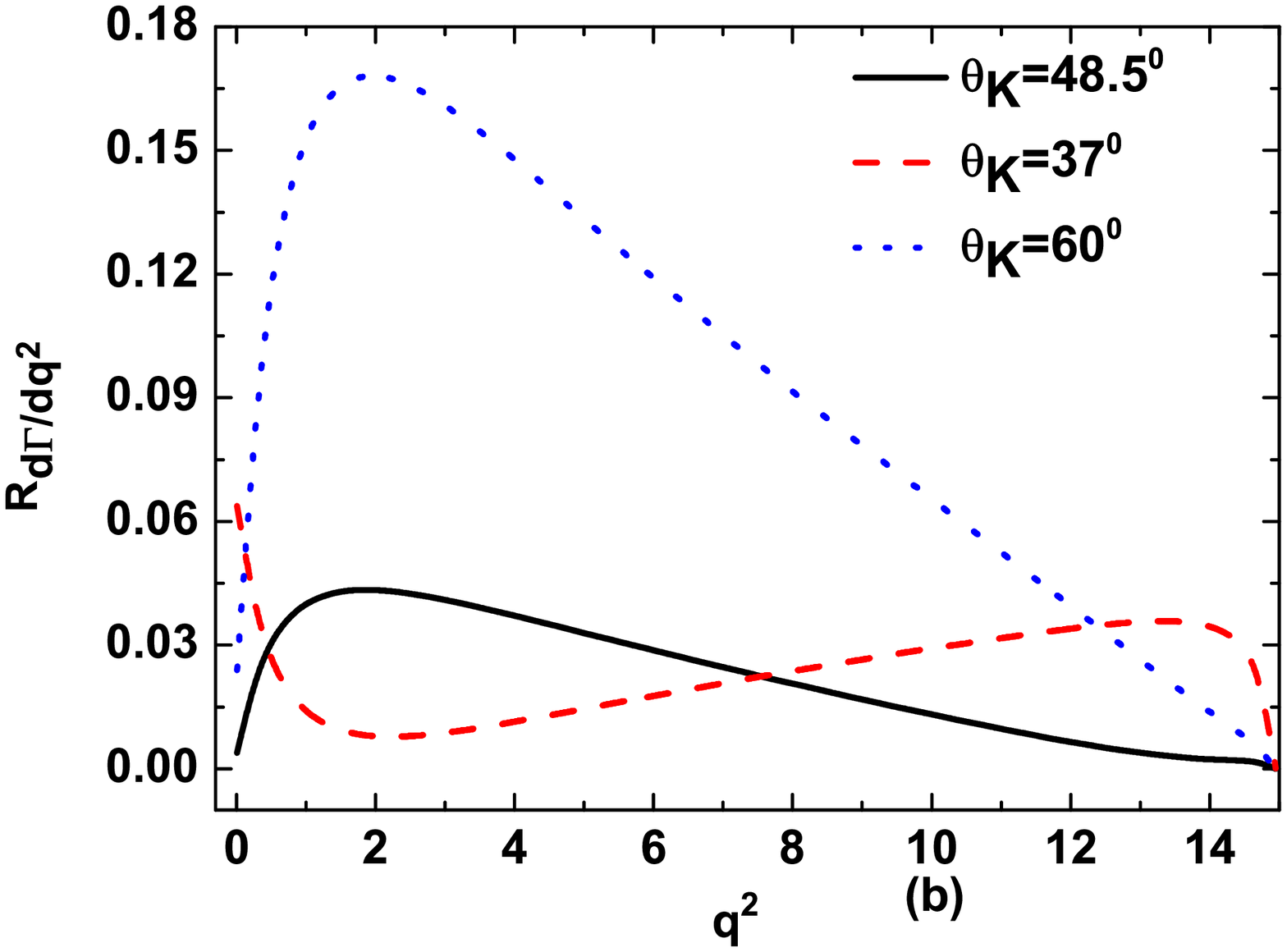}
\caption{The $q^2$ dependence of $R_{d\Gamma/dq^2}$ without
resonances. Different lines correspond to different mixing angles.}
 \label{fig:Rq2}
 \end{center}
 \end{figure}

To shed more light on the mixing angle $\theta_K$, it is useful to
define the ratio of the partial decay widths
 \begin{eqnarray}
 R_{d\Gamma/dq^2}=\frac{d\Gamma(B\to K_1(1400)l^+l^-)/dq^2}{d\Gamma(B\to
 K_1(1270)l^+l^-)/dq^2}
 \end{eqnarray}
as a function of $q^2$. Our results for the $R_{d\Gamma/dq^2}$ are
shown in Fig. \ref{fig:Rq2}, where the resonant contributions are
not taken into account. In each of the diagrams, three lines
corresponding to different values of mixing angles are presented to
show the $\theta_K$ dependence of $R_{d\Gamma/dq^2}$. As indicated
from the two diagrams in this figure, the shape of the ratio
$R_{d\Gamma/dq^2}$ strongly depends on the mixing angle $\theta_K$.
For negative values of $\theta_K$, these is a peak in the region
$14\rm{GeV}^2 \lesssim q^2 \lesssim 15\rm{GeV}^2$ and it becomes
shaper when $\theta_K$ decreases from $-27^{\circ}$ to roughly
$-40^{\circ}$.  When $\theta_K \lesssim -48^{\circ}$, a maximum
appears at the point around $q^2=2\rm{GeV}^2$.  The situation is
also similar for the positive values of the mixing angle $\theta_K$.
These behaviors arise from the fact that the mixing angle is close
to $\pm 45^\circ$ and $B\to A$ form factors have similar magnitudes.

\begin{figure}
 \begin{center}
 \includegraphics[width=7.cm]{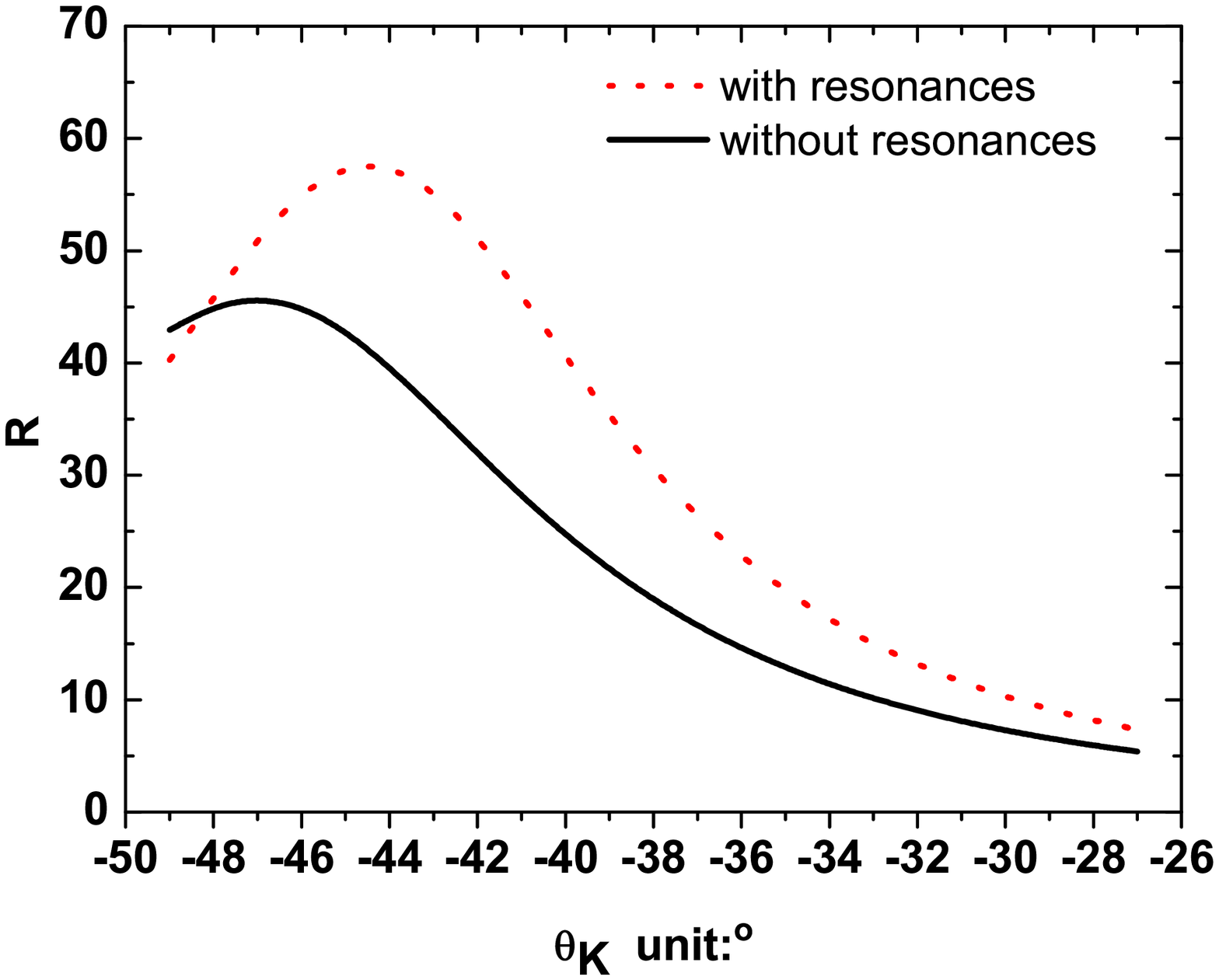}
 \includegraphics[width=7.cm]{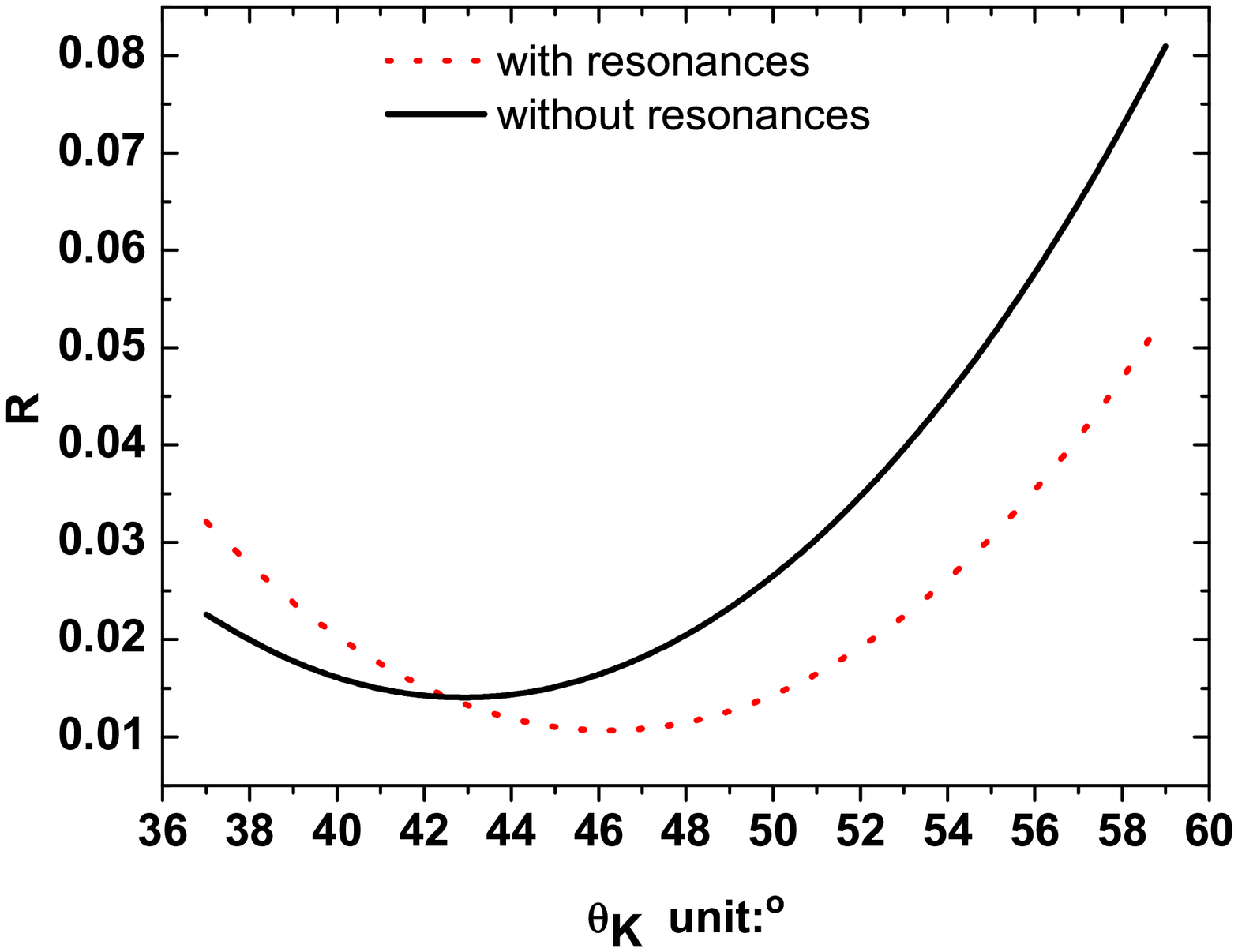}
 \caption{The $\theta_K$ dependence of $R$. The black
solid(red dot) line represents the results without(with) the
resonant contributions.}
 \label{fig:Rtheta}
 \end{center}
 \end{figure}

We also calculate the ratio
 \begin{eqnarray}
 R=\frac{{\cal BR}(B\to K_1(1400)l^+l^-)}{{\cal BR}(B\to K_1(1270)l^+l^-)},
 \end{eqnarray}
and the dependence on $\theta_K$ is depicted in
Fig.~\ref{fig:Rtheta}. From these two diagrams, we can see that the
ratio $R$ is very sensitive to the value of mixing angle. The
maximum($R_{\rm{max}}\approx 46$) appears roughly at
$\theta_K=-47^{\circ}$, while the minimum($R_{\rm{min}}\approx
0.015$) appears at $\theta_K=43^{\circ}$. If we adopt the two
reference points, the predictions on $R$ are given by
\begin{eqnarray}
 R=19_{-18}^{+73},\;\;\; R=0.02_{-0.01}^{+0.07}.
\end{eqnarray}

\subsection{Forward-backward asymmetry}

The differential  forward-backward asymmetry of $\bar B\to \bar
K_1l^+l^-$ is defined by
\begin{eqnarray}
 \frac{d A_{FB}}{dq^2}&=&{\int_0^1 d\cos\theta_1 \frac{d^2\Gamma}{dq^2 d\cos\theta_1}
 -\int_{-1}^0 d\cos\theta_1 \frac{d^2\Gamma}{dq^2 d\cos\theta_1}},
\end{eqnarray}
while  the normalized differential  forward-backward asymmetry is
defined by
\begin{eqnarray}
 \frac{d \bar A_{FB}}{dq^2}&=&\frac{\frac{d  A_{FB}}{dq^2}}
 { \frac{d\Gamma}{dq^2}}=\frac{3}{4}\times \frac{-|H(L,+)|^2+|H(R,+)|^2+|H(L,-)|^2-|H(R,-)|^2}
 {|H(L,0)|^2+|H(R,0)|^2+|H(L,+)|^2+|H(R,+)|^2+|H(L,-)|^2+|H(R,-)|^2}.
\end{eqnarray}

 \begin{figure}
 \begin{center}
 \includegraphics[width=7.cm]{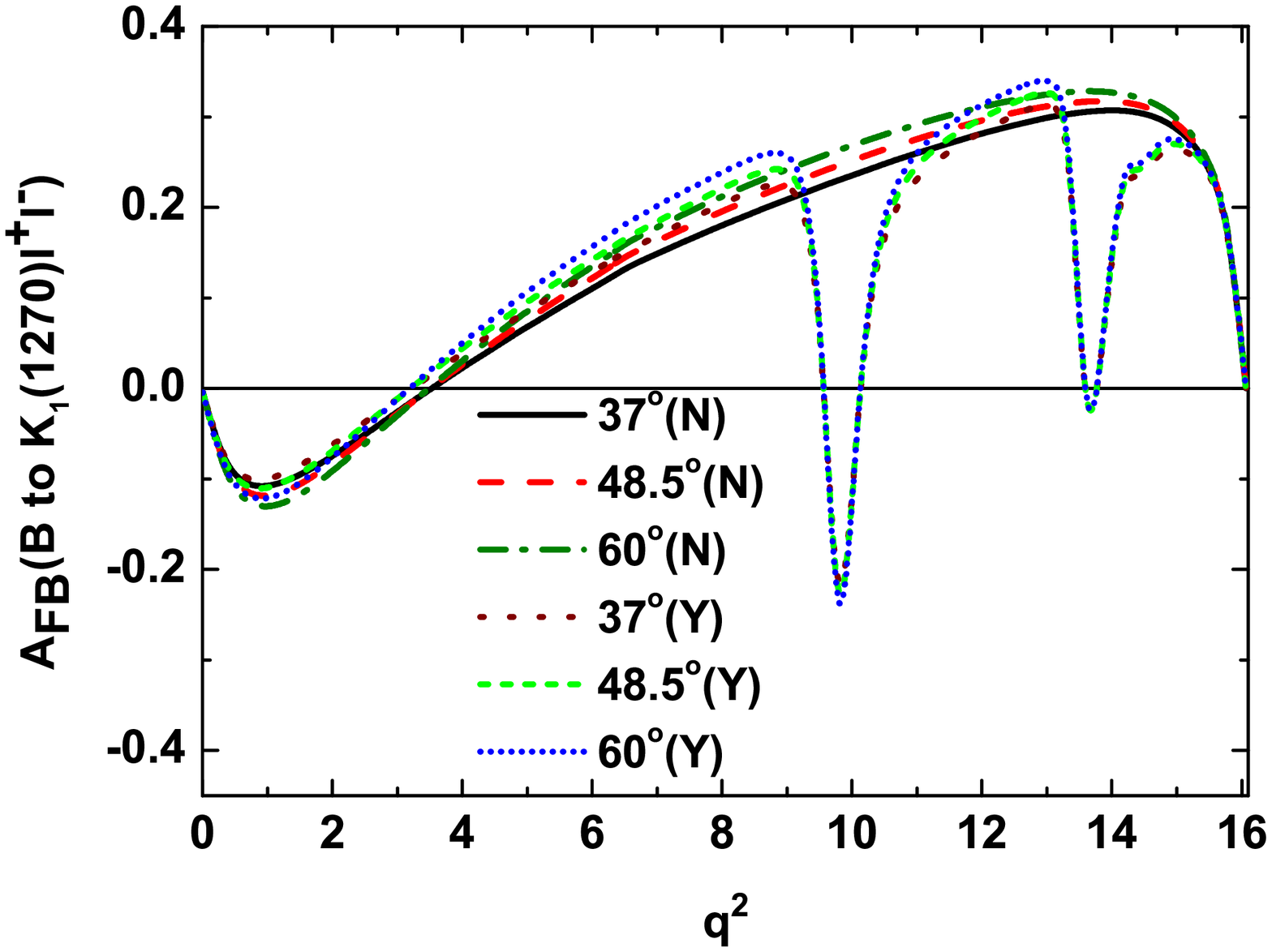}
 \includegraphics[width=7.cm]{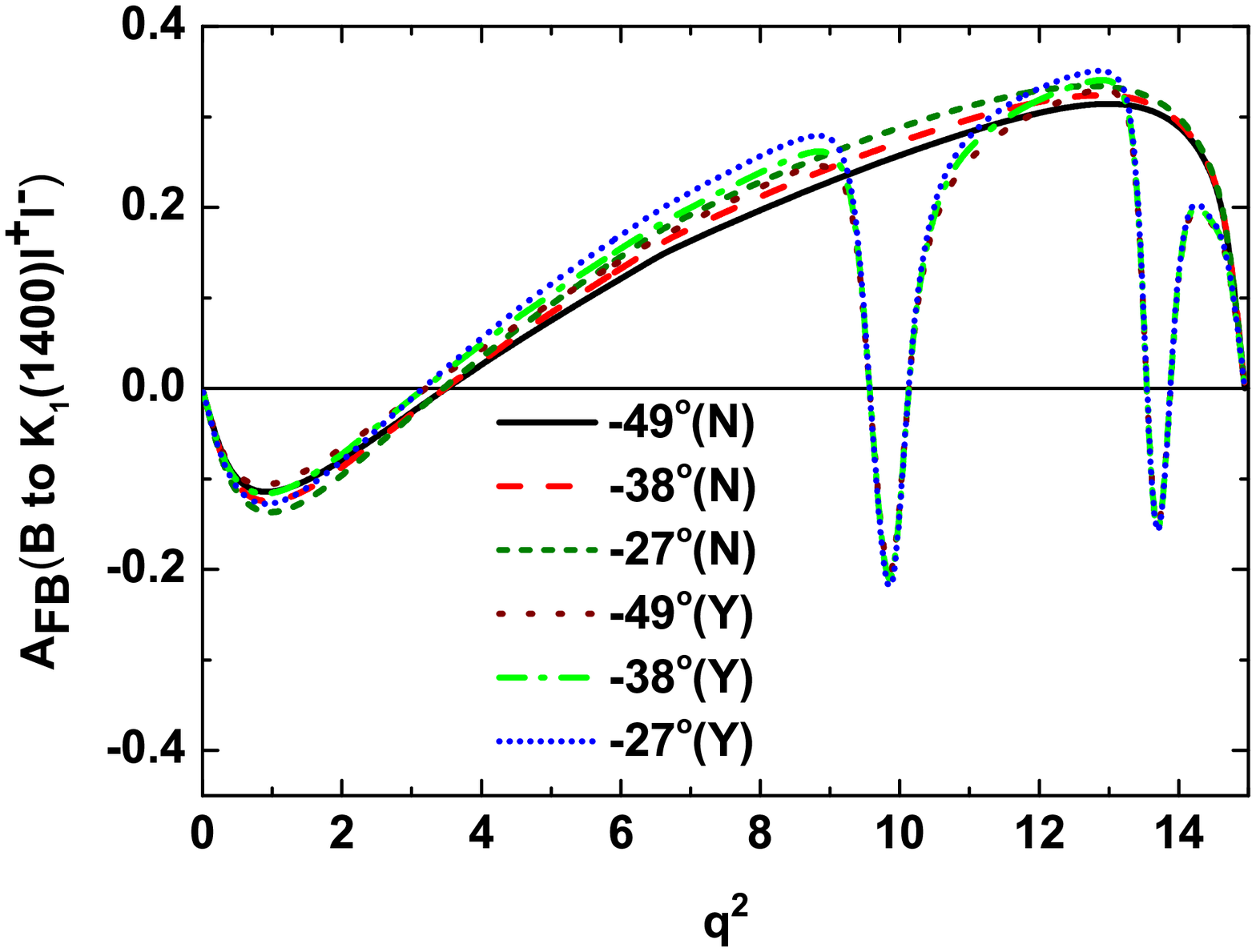}
\caption{Dependence on the $q^2$ of the normalized forward-backward
asymmetry $\frac{d \bar A_{FB}}{dq^2}$. Results corresponding to
different mixing angles are given. The character "Y"("N") in the
brackets in the diagrams denotes that the resonant contributions are
(not) taken into account.}
 \label{fig:FBSq2}
 \end{center}
 \end{figure}

When $\theta_K=-38^{\circ}[48.5^{\circ}]$, form factors of $B\to
K_{1B}$ and $B\to K_{1A}$ give destructive contributions to $B \to
K_1(1270)l^+l^-$[$B \to K_1(1400)l^+l^-$] decay. Thus the results
suffer from large uncertainties. So in Fig.~\ref{fig:FBSq2}, we only
give our results on the normalized forward-backward asymmetries
$\frac{d \bar A_{FB}}{dq^2}$ for $B \to K_1(1270)l^+l^-$ with
$\theta_K=48.5^{\circ}$ and $B \to K_1(1400)l^+l^-$ with
$\theta_K=-38^{\circ}$. From these diagrams, one can see that the
differential  forward-backward asymmetries are not sensitive to the
mixing angles. When the resonant contributions are absent, the
positions of zeros of the forward-backward asymmetries are roughly
$q^2=3.55\rm{GeV}$ for both diagrams.

\subsection{Angular distributions of $\bar B\to \bar K_1l^+l^-\to
\bar K\pi\pi l^+l^-$}

Experimentally, the $K_1$ meson is reconstructed by at least three
pseudoscalar mesons, thus the cascade decay $\bar B\to \bar
K_1l^+l^-\to \bar K\pi\pi l^+l^-$, rather than  $\bar B\to \bar
K_1l^+l^-$, will be observed. For the three-body decay $K_1\to
K\pi\pi$, there is an additional phenomenological amplitude $F_m$,
where $m$ denotes the spin eigenvalue of $K_1$ along the normal
direction to the decay plane. For $\bar B\to \bar K_1l^+l^-\to (\bar
K\pi\pi)l^+l^-$, both of helicity $\lambda$ and spin eigenvalue $m$
run from $-1$ to 1. When integrating over the rotation angle around
the normal to the decay plane, the interference between different
$F_m$ vanishes and this gives $2J+1$ real parameters
$|R_m|^2$~\cite{Berman:1965gi}. Parity conservation provides
additional constraints: only the $m$ satisfying the relation
$P=(-1)^{m+1}$ contributes. To be more specific, $m=\pm1$ is
required for the $\bar B\to \bar K_1l^+l^-\to \bar K\pi\pi l^+l^-$
transition since the quantum numbers for $K_1$ are $J^P=1^+$.

Following the angular distributions of $B\to VA$
decays~\cite{Datta:2007yk},  the angular distributions of $\bar B\to
\bar K_1l^+l^-\to\bar  K\pi\pi l^+l^-$ are derived  as
\begin{eqnarray}
  \frac{d^4\Gamma}{dq^2d\cos\theta_1d\cos\theta_2d\phi}&\propto& \frac{1}{\sum_m|R_m|^2}\sum_m|R_m|^2
   \left[\left|\sum_\lambda H(L,\lambda) L(L,\lambda)
  d_{\lambda, m}^{J_2}(\theta_2)\right|^2 +\left|\sum_\lambda H(R,\lambda) L(R,\lambda)
  d_{\lambda, m}^{J_2}(\theta_2)\right|^2\right]\nonumber\\
  &=& 2q^2\bigg\{  2\big(|H(L,0)|^2+|H(R,0)|^2)\big) \sin^2\theta_1 \sin^2\theta_2\nonumber\\
  &&+\frac{|H(L,1)|^2+|H(R,-1)|^2}{2}(1-\cos\theta_1)^2(1+\cos^2\theta_2)\nonumber\\
  &&+\frac{|H(L,-1)|^2+|H(R,1)|^2}{2}(1+\cos\theta_1)^2(1+\cos^2\theta_2)\nonumber\\
  &&+{\rm  Re}\big[H(L,0)H(L,1)^*+H(R,0)H(R,-1)^*\big](1-\cos\theta_1)(1+\cos\theta_2)\sin\theta_1\sin\theta_2\cos\phi  \nonumber\\
  &&+{\rm  Im}\big[H(L,0)H(L,1)^*+H(R,0)^*H(R,-1)\big](1-\cos\theta_1)(1+\cos\theta_2)\sin\theta_1\sin\theta_2\sin\phi  \nonumber\\
  &&+{\rm  Re}\big[H(L,0)H(L,-1)^*+H(R,0)H(R,1)^*\big](1+\cos\theta_1)(1-\cos\theta_2)\sin\theta_1\sin\theta_2\cos\phi  \nonumber\\
  &&+{\rm  Im}\big[H(L,0)^*H(L,-1)+H(R,0)H(R,1)^*\big](1+\cos\theta_1)(1-\cos\theta_2)\sin\theta_1\sin\theta_2\sin\phi  \nonumber\\
  &&+{\rm  Re}\big[H(L,1)H(L,-1)^*+H(R,1)H(R,-1)^*\big]\sin^2\theta_1\sin^2\theta_2\cos(2\phi) \nonumber\\
  &&+{\rm  Im}\big[H(L,1)^*H(L,-1)+H(R,1)^*H(R,-1)\big]\sin^2\theta_1\sin^2\theta_2\sin(2\phi) \nonumber\\
  &&-{\rm  Re}\big[H(R,0)H(R,1)^*+H(L,0)H(L,-1)^*\big](1+\cos\theta_1)(1+\cos\theta_2)\sin\theta_1\sin\theta_2\cos\phi  \nonumber\\
  &&-{\rm  Im}\big[H(R,0)H(R,1)^*+H(L,0)^*H(L,-1)\big](1+\cos\theta_1)(1+\cos\theta_2)\sin\theta_1\sin\theta_2\sin\phi  \nonumber\\
  &&-{\rm  Re}\big[H(R,0)H(R,-1)^*+H(L,0)H(L,1)^*\big](1-\cos\theta_1)(1-\cos\theta_2)\sin\theta_1\sin\theta_2\cos\phi  \nonumber\\
  &&-{\rm  Im}\big[H(R,0)^*H(R,-1)+H(L,0)H(L,1)^*\big](1-\cos\theta_1)(1-\cos\theta_2)\sin\theta_1\sin\theta_2\sin\phi \bigg\}
 \nonumber\\
 &&+2q^2r_1\bigg\{({|H(L,1)|^2-|H(R,-1)|^2})(1-\cos\theta_1)^2 \cos\theta_2 \nonumber\\
  &&+({|H(R,1)|^2-|H(L,-1)|^2})(1+\cos\theta_1)^2 \cos\theta_2\nonumber\\
  &&+{\rm  Re}\big[H(L,0)H(L,1)^*-H(R,0)H(R,-1)^*\big](1-\cos\theta_1)(1+\cos\theta_2)\sin\theta_1\sin\theta_2\cos\phi  \nonumber\\
  &&+{\rm  Im}\big[H(L,0)H(L,1)^*-H(R,0)^*H(R,-1)\big](1-\cos\theta_1)(1+\cos\theta_2)\sin\theta_1\sin\theta_2\sin\phi  \nonumber\\
  &&+{\rm  Re}\big[H(L,0)H(L,-1)^*-H(R,0)H(R,1)^*\big](1+\cos\theta_1)(1-\cos\theta_2)\sin\theta_1\sin\theta_2\cos\phi  \nonumber\\
  &&+{\rm  Im}\big[H(L,0)^*H(L,-1)-H(R,0)H(R,1)^*\big](1+\cos\theta_1)(1-\cos\theta_2)\sin\theta_1\sin\theta_2\sin\phi  \nonumber\\
  &&-{\rm  Re}\big[H(R,0)H(R,1)^*-H(L,0)H(L,-1)^*\big](1+\cos\theta_1)(1+\cos\theta_2)\sin\theta_1\sin\theta_2\cos\phi  \nonumber\\
  &&-{\rm  Im}\big[H(R,0)H(R,1)^*-H(L,0)^*H(L,-1)\big](1+\cos\theta_1)(1+\cos\theta_2)\sin\theta_1\sin\theta_2\sin\phi  \nonumber\\
  &&-{\rm  Re}\big[H(R,0)H(R,-1)^*- H(L,0)H(L,1)^*\big](1-\cos\theta_1)(1-\cos\theta_2)\sin\theta_1\sin\theta_2\cos\phi  \nonumber\\
  &&-{\rm  Im}\big[H(R,0)^*H(R,-1)-H(L,0)H(L,1)^*\big](1-\cos\theta_1)(1-\cos\theta_2)\sin\theta_1\sin\theta_2\sin\phi   \bigg\},
\end{eqnarray}
where 
%
the asymmetry parameter $r_1$ is defined as
\begin{eqnarray}
 r_1=\frac{|R_1|^2-|R_{-1}|^2}{|R_1|^2+|R_{-1}|^2}.
\end{eqnarray}
It should be pointed out that this parameter depends on the dynamics
of $K_1\to K\pi\pi$. $r_1$ would vanish if a symmetry with respect
to the inversion of the normal to the decay plane is satisfied. For
example, if the three-body decay goes through $K_1\to K\rho\to
K\pi\pi$, the parameter $r_1$ is zero.  The branching ratio of
$K_1(1270)\to \rho K$ is very large, thus we expect that the terms
proportional to $r_1$ will not contribute a lot to the $B\to
K_1(1270)l^+l^-\to K\pi\pi l^+l^-$ channel. But for the $K_1(1400)$
meson, the dominant channel is $K_1(1400)\to K^*\pi$ and this kind
of symmetry does not exist.

\section{Conclusions}

Decays induced by FCNC have typically tiny branching fractions in
the SM, which are very sensitive to the NP scenarios. Semileptonic
$B\to K^*l^+l^-$ and $B\to K_1l^+l^-$ decays are ideal probes to
detect the NP effect, as their observables receive less
nonperturbative pollution than nonleptonic $B$ decays.

Using the $B\to K_1$ form factors evaluated in the PQCD approach, we
study the semileptonic $B\to K_1(1270)l^+l^-$ and $B\to
K_1(1400)l^+l^-$ decays, where $l=e,\mu$. Applying the technique of
helicity amplitudes, we express decay amplitudes in terms of several
independent and Lorentz invariant pieces. We study the total
branching fractions and polarizations of $\bar B\to \bar K_1l^+l^-$
decays. $K_1(1270)$ and $K_1(1400)$ are mixtures of $K_{1A}$ and
$K_{1B}$ which are $^3P_1$ and $^1P_1$ states, respectively. The
ambiguity in the sign of the mixing angle will induce much large
differences to branching ratios of semileptonic $B$ decays:
branching ratios without resonant contributions either have the
order of $10^{-6}$ or $10^{-8}$. The future measurements of the
branching fractions are helpful to discriminate the internal
structures of the two strange mesons.  Large differences in
polarizations are also produced by the different mixing angles. We
show that the long-distance contributions will sizably change the
dilepton invariant mass distributions in the resonant region. Since
the $K_1$ meson can not be directly detected, experimentalists can
perform the angular distribution analysis for the $ \bar B\to \bar
K_1l^+l^-\to (\bar K\pi\pi)l^+l^-$ decay channels which contain more
information on the internal structures. With the help of helicity
amplitudes, we directly give these angular distributions of the
$\bar B\to \bar K_1l^+l^-\to (\bar K\pi\pi)l^+l^-$ decays.

\section*{Acknowledgements}

This work is partly supported by National Natural Science Foundation
of China under Grants No. 10735080, No. 10625525, and No. 10525523.
W. Wang would like to acknowledge Y. Jia and H.B. Li for fruitful
discussions.

\begin{appendix}

\section{Expressions for $C_9^{eff}(q^2)$}
 \label{appendix:C9eff}
$C_9^{\rm{eff}}$ in Eq.~(\ref{eq:Ampbtos}) contains both the
long-distance and short-distance contributions, which is given by
 \begin{eqnarray}
 C_9^{\rm{eff}}(q^2)&=&C_9(\mu)+Y_{\rm{pert}}(\hat{s})+Y_{\rm{LD}}(q^2).
 \end{eqnarray}
with $\hat{s}=q^2/m_B^2$. $Y_{\rm{pert}}$ represents the
perturbative contributions, and $Y_{\rm{LD}}$ is the long-distance
part. The $Y_{\rm{pert}}$ is given by~\cite{Buras:1994dj}
 \begin{eqnarray}
 Y_{\rm{pert}}(\hat{s})&=&
 h(\hat{m_c},\hat{s})C_0-\frac{1}{2}h(1,\hat{s})(4C_3+4
 C_4+3C_5+C_6)\nonumber\\
 &&-\frac{1}{2}h(0,\hat{s})(C_3+3
 C_4) + \frac{2}{9}(3C_3 + C_4 +3C_5+ C_6),\label{eq:ypert}
 \end{eqnarray}
with $C_0=C_1+3C_2+3C_3+C_4+3C_5 +C_6$ and $\hat{m}_c=m_c/m_b$. The
relevant Wilson coefficients, listed in Table \ref{tab:wilsons}, are
given up to the leading logarithmic accuracy~\cite{Buchalla:1995vs}.
The long-distance part $Y_{\rm{LD}}$ denotes the contributions of
$B\to K_1V$ resonances, where $V$ is a vector meson. Because of the
large decay width of $B\to K_1V(\bar cc)$, only the contributions
from charmonium states are taken into account\cite{Lu_and_zhang}:
 \begin{eqnarray}
 Y_{\rm{LD}}(q^2)=-\frac{3\pi}{\alpha_{\rm{em}}^2}C_0
 \sum_{V=J/\Psi...}\kappa_V\frac{m_V {\cal B}(V\to l^+l^-)\Gamma_{\rm{tot}}^V}
 {q^2-m_V^2+im_V\Gamma_{\rm{tot}}^V}.\label{eq:yld}
 \end{eqnarray}
$\kappa_V$ is introduced to give correct predictions on the decay
rates of $B\to K_1V(c\bar c)$ in the factorization approach. With
the available data, this parameter can be obtained through fitting
the decay rates. For example, $\kappa_V$ for $B\to J/\Psi K^*$ is
determined as $\kappa_V=2.3$~\cite{Ligeti:1995yz}. Except for the
branching ratio of $\bar B^0\to J/\Psi \bar
K_1(1270)$~\cite{Amsler:2008zzb}, there is no experimental study on
$B\to K_1V(c\bar c)$. We will assume the same value for $\kappa_V$
in $B\to K_1V(c\bar c)$ due to the lack of data.  In Table
\ref{tab:charmonium}, we list the properties of the vector
charmonium states: mass, width, and branching fractions of the
leptonic decay channel $V\to l^+l^-$~~\cite{Amsler:2008zzb}.

 \begin{table}
 \caption{The values of Wilson coefficients $C_i(m_b)$ in the leading
logarithmic approximation, with $m_W=80.4\mbox{GeV}$, $\mu=m_{b,\rm
pole}$~\cite{Buchalla:1995vs}.}
 \label{tab:wilsons}
 \begin{center}
 \begin{tabular}{c c c c c c c c c}
 \hline\hline
 \ \ \ $C_1$ &$C_2$ &$C_3$ &$C_4$ &$C_5$ &$C_6$ &$C_7^{\rm{eff}}$ &$C_9$ &$C_{10}$       \\
 \ \ \ $1.107$   &$-0.248$   &$-0.011$    &$-0.026$    &$-0.007$    &$-0.031$    &$-0.313$    &$4.344$    &$-4.669$    \\
 \hline\hline
 \end{tabular}
 \end{center}
 \end{table}

\begin{table}
\caption{Masses,  decay widths and branching fractions of dilepton
decays of vector charmonium
states~\cite{Amsler:2008zzb}.}\label{tab:charmonium}
\begin{center}
\begin{ruledtabular}
\begin{tabular}{cccc}
$V$ & Mass[\rm{GeV}] &  $\Gamma_{\rm tot}^V$[\rm{MeV}]
 &${\cal BR}(V\to l^+ l^-)$ with $l=e,\mu$
\\
\hline $J/\Psi$ & $3.097$ & $0.093$ & $5.9\times10^{-2}$
\\
$\Psi(2S)$   & $3.686$ & $0.327$ & $7.4\times10^{-3}$
\\
$\Psi(3770)$ & $3.772$ & $25.2$ & $9.8\times10^{-6}$
\\
$\Psi(4040)$ & $4.040$ & $80$ & $1.1\times10^{-5}$
\\
$\Psi(4160)$ & $4.153$ & $103$ & $8.1\times10^{-6}$
\\
$\Psi(4415)$ & $4.421$ & $62$ & $9.4\times10^{-6}$
\end{tabular}
\end{ruledtabular}
\end{center}
\end{table}

\section{Mixing between $K_1(1270)$ and $K_1(1400)$}
\label{appendix:mixing}

The physical states $K_1(1270)$ and $K_1(1400)$ are mixtures of the
$K_{1A}$ and $K_{1B}$ states with the mixing angle $\theta_K$:
\begin{eqnarray}
|K_1(1270)\rangle&=&|K_{1A}\rangle
{\rm{sin}}\theta_K+|K_{1B}\rangle{\rm{cos}}\theta_K,\\
|K_1(1400)\rangle&=&|K_{1A}\rangle
{\rm{cos}}\theta_K-|K_{1B}\rangle{\rm{sin}}\theta_K.
\end{eqnarray}
In the flavor SU(3) symmetry limit, these mesons do not mix with
each other; but since the $s$ quark is heavier than the $u,d$
quarks, $K_1(1270)$ and $K_1(1400)$ are not purely $1^3P_1$ or
$1^1P_1$ states. Generally, the mixing angle can be determined by
the experimental data. One feasible method is making use of the
decay $\tau^-\to K_1\nu_\tau$, whose partial decay rate is given by
\begin{eqnarray}
{\Gamma}(\tau^-\to
K_1\nu_\tau)=\frac{m_\tau^3}{16\pi}G_F^2|V_{us}|^2f_A^2\left(1-\frac{m_A^2}{m_\tau^2}
\right)^2\left(1+\frac{2m_A^2}{m_\tau^2}\right),
\end{eqnarray}
with the measured results for branching
fractions~\cite{Amsler:2008zzb}
\begin{eqnarray}
{\cal BR}(\tau^-\to K_1(1270)\nu_\tau)=(4.7\pm1.1)\times 10^{-3},\;
{\cal BR}(\tau^-\to K_1(1400)\nu_\tau)=(1.7\pm2.6)\times
10^{-3}.\label{eq:BRintau}
\end{eqnarray}
The longitudinal decay constants (in MeV) can be straightforwardly
obtained
\begin{eqnarray}
|f_{K_1(1270)}|=169_{-21}^{+19};\;\;\;
|f_{K_1(1400)}|=125_{-125}^{+~74}.\label{eq:decayconstantK1}
\end{eqnarray}
In principle, one can combine the decay constants for $K_{1A}$,
$K_{1B}$ evaluated in QCD sum rules~\cite{Yang:2007zt} with the
above results to determine the mixing angle $\theta_K$. But since
there are large uncertainties in Eq.~(\ref{eq:decayconstantK1}), the
constraint on the mixing angle is expected to be rather smooth:
\begin{eqnarray}
 -49^\circ<\theta_K<-27^\circ,\;\;\;{\rm or}\;\;
 37^\circ<\theta_K<60^\circ,\label{eq:mixingangleresults}
\end{eqnarray}
where we have taken the uncertainties from the branching ratios in
Eq.(\ref{eq:BRintau}) and the first Gegenbauer moment $a_1^{K_1}$
into account but neglected the mass differences. As indicated from
Eq.~(\ref{eq:mixingangleresults}), the mixing angle $\theta_K$ still
has large uncertainties. To reduce the uncertainties, we have
proposed to use $\bar B^0\to D^+K_1^-$ to constrain the mixing
angles~\cite{Li:2009tx}. At present, we will use the two reference
points:
\begin{eqnarray}
 \theta_K=(-38\pm11)^\circ,\;\;\; \theta_K=(48.5\pm11.5)^\circ
\end{eqnarray}

\section{Helicity amplitudes}
\label{appendix:amplitudes}
Decay amplitudes for $b\to sl^+l^-$ decays in Eq.~(\ref{eq:Ampbtos})
can be rearranged as
\begin{eqnarray}
 {\cal A}(b\to sl^+l^-)&=& \frac{G_F}{\sqrt
 2}\frac{\alpha_{em}}{\pi} V_{tb}V_{ts}^* \left( \frac{C_9^{\rm eff}+C_{10}}{4}[\bar sb]_{V-A} [\bar ll]_{V+A}
 + \frac{C_9^{\rm eff}-C_{10}}{4}[\bar sb]_{V-A} [\bar ll]_{V-A}\right. \nonumber\\
 && \left. -C_{7L} m_b[\bar s i\sigma_{\mu\nu}(1+\gamma_5) b]\frac{q^\nu}{q^2}[\bar l\gamma^\mu l]
 -C_{7R} m_b[\bar s i\sigma_{\mu\nu}(1-\gamma_5) b]\frac{q^\nu}{q^2}[\bar l\gamma^\mu
 l]\right),
\end{eqnarray}
where $C_{7L}=\frac{m_b}{m_B}C_7^{\rm eff}$ and
$C_{7R}=\frac{m_s}{m_B}C_7^{\rm eff}$. The decay amplitudes for the
hadronic $\bar B\to \bar K_1l^+l^-$ decays can be obtained by
replacing the hadronic spinors $[\bar sb]$ by the $B\to K_1$ form
factors which are defined in Eq.~\eqref{eq:BtoK1formfactor}. To
predict physical observables such as partial decay widths, one needs
to evaluate the amplitude square together with the phase space.
Under the summation of different spins, spinors and polarization
vectors can be simplified. But we can see that there are still six
form factors contributing to the $B\to K_1l^+l^-$ decays,  even if
the lepton's masses are neglected. In the following, we will use a
rather simple way to derive the decay amplitudes of $B\to
K_1l^+l^-$: the helicity amplitudes.

Suppose there exists an intermediate vector state whose momentum is
denoted as $q$. The polarization vectors are denoted as
$\epsilon(\lambda)$, where $\lambda=0,\pm$ denotes the three kinds
of polarizations.  The metric tensor $g_{\mu\nu}$ can be decomposed
into  combinations of polarization vectors and momentum:
\begin{eqnarray}
 g_{\mu\nu}=-\sum_\lambda\epsilon_\mu(\lambda) \epsilon^*_\nu(\lambda) +\frac{q_\mu q_\nu
 }{q^2}.\label{eq:replacement}
\end{eqnarray}
In the SM,  the lepton pair in the final state is produced via an
off-shell photon, or a $Z$ boson or some possible hadronic vector
mesons. The different intermediate states may give different
couplings but the amplitudes share many commonalities: the Lorentz
structure for the vertex of the lepton pair  is either $V-A$ or
$V+A$ or any combinations of them. Thus the decay amplitudes of
$\bar B\to \bar K_1l^+l^-$ can be redefined as
\begin{eqnarray}
 {\cal A}(\bar B\to \bar K_1l^+l^-)&=& {L}_\mu(L) {H}_\mu(L)+{L}_\mu(R) {H}_\mu(R),
\end{eqnarray}
where ${L}(L),{L}(R)$ are the lepton pair spinor products:
\begin{eqnarray}
 {L}(L)&=& \bar l\gamma_\mu(1-\gamma_5)l,\;\;\;
 {L}(R)= \bar l\gamma_\mu(1+\gamma_5)l.
\end{eqnarray}
Inserting a metric tensor $g_{\mu\nu}$ and substituting the identity
in Eq.~\eqref{eq:replacement} into the decay amplitudes, one obtain
two independent parts:
\begin{eqnarray}
 {\cal A}(\bar B\to \bar K_1l^+l^-)&=& {L}_\mu(L) {H}_\nu(L) g^{\mu\nu}+{L}_\mu(R) {H}_\nu(R)
 g^{\mu\nu}\nonumber\\
  &=&-\sum_{\lambda}{L}(L,\lambda) {H}(L,\lambda)  -\sum_{\lambda}{L}(R,\lambda) {
  H}(R,\lambda),
\end{eqnarray}
where $q^\mu$ is the momentum of the lepton pair.  ${
L}(L,\lambda)={L}^\mu(L) \epsilon_\mu(\lambda)$ and ${
L}(R,\lambda)={L}^\mu(R) \epsilon_\mu(\lambda)$ denote the Lorentz
invariant amplitudes for the lepton part. It is also similar for the
Lorentz invariant hadronic amplitudes: ${ H}(L,\lambda)={ H}^\mu(L)
\epsilon^*_\mu(\lambda)$ and ${ H}(R,\lambda)={ H}^\mu(R)
\epsilon^*_\mu(\lambda)$. The last term (proportional to $q^\mu$) in
the metric tensor $g_{\mu\nu}$ vanishes: using equation of motion,
this term is proportional to the mass of the lepton which has been
set to zero.

An advantage of the helicity amplitudes is that both of the hadronic
amplitudes and the leptonic amplitudes are Lorentz invariant. So one
can choose any convenient frame to evaluate them separately. Usually
the lepton amplitudes are evaluated in the central mass frame of the
lepton pair, while the hadronic decay amplitudes can be directly
obtained in the $B$ meson rest frame. The phase space for multibody
decays can also be written into several Lorentz invariant pieces. We
will take the three-body decays' phase space as an example:
\begin{eqnarray}
 d\Phi_3(P;p_1,p_2,p_3)&=&\int \frac{d^3p_1}{(2\pi)^32E_1} \frac{d^3p_2}{(2\pi)^32E_2}
 \frac{d^3p_3}{(2\pi)^32E_3}\delta^4(P-p_1-p_2-p_3)
\end{eqnarray}
which can be rearranged as
\begin{eqnarray}
 d\Phi_3(P;p_1,p_2,p_3)&=&d\Phi_2(q,p_1,p_2)\times
 d\Phi_2(P;q,p_3)\times (2\pi)^3 dq^2.
\end{eqnarray}

\begin{figure}
\begin{center}
\includegraphics[scale=0.5]{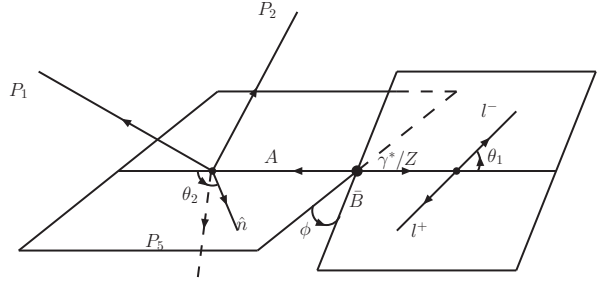}
\end{center}
\caption{Definitions of helicity angles $(\theta_1,\theta_2,\phi)$
in $\bar B\to \bar K_1l^+l^-\to K\pi\pi l^+l^-$ decay. The direction
$\hat n$ is the normal to the decay plane of $K_1$.  The angle
between the direction $\hat n$ and the $K_1$ moving direction in the
$B$ meson rest frame is defined as $\theta_2$, while the angle
between the moving direction of $l^-$ in the lepton pair (a
$\gamma^*$ or a $Z$ boson)  rest frame and the moving direction of
the lepton pair in the $B$ meson rest frame is defined as
$\theta_1$. The angle, between the $l^+l^-$ decay plane and the
plane defined by the $K_1$ moving direction in the $B$ meson rest
frame and $\hat n$, is defined as $\phi$. } \label{Feyn:angular}
\end{figure}

Before the results for these amplitudes are presented, we will give
our convention on the helicity angles which are depicted in
Fig.~\ref{Feyn:angular}. Experimentally, the $K_1$ meson will be
reconstructed by the three pseudoscalar state $K\pi\pi$. In the rest
frame of the $K_1$ meson, the final three mesons move in one decay
plane and the direction normal to this decay plane is denoted as
$\hat n$. The angle between the direction $\hat n$ and the $K_1$
moving direction in the $B$ meson rest frame is defined as
$\theta_2$, while the angle between the moving direction of $l^-$ in
the lepton pair rest frame and the moving direction of the lepton
pair in the $B$ meson rest frame is defined as $\theta_1$. The angle
between the $l^+l^-$ decay plane and the plane defined  by the $K_1$
moving direction in the $B$ meson rest frame and $\hat n$ is defined
as $\phi$.

In the rest frame of the lepton pair,  the leptonic decay amplitudes
are evaluated as
\begin{eqnarray}
 {L}(L,0)&=& 2\sqrt {q^2}\sin\theta_1,\;\;\;\\
 {L}(L,+)&=& -2\sqrt 2\sqrt {q^2}\sin^2\frac{\theta_1}{2}
 e^{i\phi},\;\;\;\\
 {L}(L,-)&=& -2\sqrt 2\sqrt {q^2}\cos^2\frac{\theta_1}{2}
 e^{-i\phi},\\
 {L}(R,0)&=& -2\sqrt {q^2}\sin\theta_1,\;\;\;\\
 {L}(R,+)&=& -2\sqrt 2\sqrt {q^2}\cos^2\frac{\theta_1}{2}
 e^{i\phi},\;\;\;\\
 {L}(R,-)&=& -2\sqrt 2\sqrt {q^2}\sin^2\frac{\theta_1}{2}
 e^{-i\phi}.
\end{eqnarray}
%
In the $B$ meson rest frame,  the hadronic transition amplitudes are
given by
\begin{eqnarray}
 { H}(L,0)&=&\frac{G_F V_{tb} V^*_{ts} \alpha_{em}}{8\sqrt{2} \pi m_A \sqrt{q^2}}
 \bigg\{2(C_{7L}+C_{7R})m_b\left[\frac{\lambda T_3(q^2)}{m_B^2-m_A^2}
 -\left(3m_A^2+m_B^2-q^2\right)T_2(q^2)\right]\nonumber\\
 &&+(C_9^{\rm eff}-C_{10})\left[(m_B-m_A)(m_A^2-m_B^2+q^2)V_1(q^2)
 +\frac{\lambda V_2(q^2)}{(m_B-m_A)}\right]\bigg\},\;\;\;\label{eq:HL0}
\end{eqnarray}
\begin{eqnarray}
 {H}(L,+)&=&\frac{G_F V_{tb} V^*_{ts}\alpha_{em}}{4\sqrt{2}\pi q^2}
 \bigg\{2(C_{7L}-C_{7R})m_b\sqrt{\lambda} T_1(q^2) - 2(C_{7L}+C_{7R}) m_b (m_B^2-m_A^2) T_2(q^2) \nonumber\\
 &&+(C_9^{\rm eff}-C_{10})q^2\bigg[\frac{\sqrt{\lambda} A(q^2)}{(m_B-m_A)}-(m_B-m_A)V_1(q^2)\bigg]\bigg\},\;\;\;\label{eq:HL+}
\end{eqnarray}
\begin{eqnarray}
 {H}(L,-)&=&\frac{G_F V_{tb} V^*_{ts}\alpha_{em}}{4\sqrt{2}\pi q^2}
 \bigg\{-2(C_{7L}-C_{7R})m_b\sqrt{\lambda} T_1(q^2) - 2(C_{7L}+C_{7R}) m_b (m_B^2-m_A^2) T_2(q^2) \nonumber\\
 &&+(C_9^{\rm eff}-C_{10})q^2\bigg[-\frac{\sqrt{\lambda} A(q^2)}{(m_B-m_A)}-(m_B-m_A)V_1(q^2)\bigg]\bigg\},\;\;\;\label{eq:HL-}
\end{eqnarray}
\begin{eqnarray}
 {H}(R,0)&=&\frac{G_F V_{tb} V^*_{ts} \alpha_{em}}{8\sqrt{2} \pi m_A \sqrt{q^2}}
 \bigg\{2(C_{7L}+C_{7R})m_b\left[\frac{\lambda T_3(q^2)}{m_B^2-m_A^2}
 -\left(3m_A^2+m_B^2-q^2\right)T_2(q^2)\right]\nonumber\\
 &&+(C_9^{\rm eff}+C_{10})\left[(m_B-m_A)(m_A^2-m_B^2+q^2)V_1(q^2)
 +\frac{\lambda V_2(q^2)}{(m_B-m_A)}\right]\bigg\},\;\;\;\label{eq:HR0}
\end{eqnarray}
\begin{eqnarray}
 {H}(R,+)&=&\frac{G_F V_{tb} V^*_{ts}\alpha_{em}}{4\sqrt{2}\pi q^2}
 \bigg\{2(C_{7L}-C_{7R})m_b\sqrt{\lambda} T_1(q^2) - 2(C_{7L}+C_{7R}) m_b (m_B^2-m_A^2) T_2(q^2) \nonumber\\
 &&+(C_9^{\rm eff}+C_{10})q^2\bigg[\frac{\sqrt{\lambda} A(q^2)}{(m_B-m_A)}-(m_B-m_A)V_1(q^2)\bigg]\bigg\},\;\;\;\label{eq:HR+}
\end{eqnarray}
\begin{eqnarray}
 {H}(R,-)&=&\frac{G_F V_{tb} V^*_{ts}\alpha_{em}}{4\sqrt{2}\pi q^2}
 \bigg\{-2(C_{7L}-C_{7R})m_b\sqrt{\lambda} T_1(q^2) - 2(C_{7L}+C_{7R}) m_b (m_B^2-m_A^2) T_2(q^2) \nonumber\\
 &&+(C_9^{\rm eff}+C_{10})q^2\bigg[-\frac{\sqrt{\lambda} A(q^2)}{(m_B-m_A)}-(m_B-m_A)V_1(q^2)\bigg]\bigg\}.\;\;\;\label{eq:HR-}
\end{eqnarray}

\end{appendix}


\end{document}